\documentclass[aps,prx,reprint,superscriptaddress]{revtex4-2}
\usepackage{amsmath,amsfonts,amssymb,bm,hyperref,graphicx,subfigure,orcidlink}
\usepackage{changes}
\usepackage[notrig]{physics}
\usepackage{CJKutf8}
\usepackage{xcolor}

\UseRawInputEncoding

\newcommand{\ud}{\mathrm{d}}

\newcommand{\ue}{\mathrm{e}}
\newcommand{\ui}{\mathrm{i}}

\begin{document}

\begin{CJK*}{UTF8}{gbsn}
\title{A Nonsecular Quantum Master Equation\\ with Detailed Balance and Second-order Error Bound}

\author{Ji-Hui Pei (裴继辉)\orcidlink{0000-0002-3466-4791}}
\email{pjh@stu.pku.edu.cn}
\affiliation{Graduate School of Chinese Academy of Engineering Physics, Beijing, 100084, China}
\affiliation{School of Physics, Peking University, Beijing, 100871, China}
\author{Jin-Fu Chen (陈劲夫)\orcidlink{0000-0002-7207-969X}}
\affiliation{Instituut-Lorentz, Universiteit Leiden, P.O. Box 9506, 2300 RA Leiden, The Netherlands}
\affiliation{$\langle aQa^L\rangle$ Applied Quantum Algorithms Leiden, The Netherlands}

\begin{abstract}

An effective theory of open quantum systems should preserve the detailed balance structure inherited from microscopic reversibility 
and remain accurate throughout relaxation. 
Meeting both requirements is particularly challenging in quantum many-body systems, 
where dense spectra can invalidate the secular approximation even at weak coupling. 
Here we construct a nonsecular quantum master equation that satisfies time-reversal Kubo-Martin-Schwinger (TR-KMS) detailed balance. 
Under suitable conditions, we prove a time-uniform \(O(\lambda^2)\) bound on the difference between the predicted and exact reduced density matrices, where \(\lambda\) is the system–bath coupling strength.
Our construction completes the Redfield generator by imposing detailed balance while matching its dissipative and coherent coefficients at resonance.
In this approach, TR-KMS detailed balance is compatible with resonance matching, whereas its counterpart without time reversal is not. 
In a benchmark against the exact microscopic dynamics of two coupled oscillators, our equation maintains \(O(\lambda^2)\) accuracy across secular and nonsecular regimes, outperforming the existing detailed-balance equations tested.
These results provide a systematic route from microscopic models to thermodynamically consistent descriptions of open quantum dynamics beyond the secular approximation.

\end{abstract}

\maketitle
\end{CJK*}

\section{Introduction}

Quantum master equations serve as an indispensable framework for understanding how open quantum systems dissipate energy, lose coherence, and relax to thermal equilibrium under the influence of an environment~\cite{FazioEtAl2025,WeimerEtAl2021}.
Deriving such an equation from microscopic system-bath dynamics is a central task in open quantum physics \cite{BreuerPetruccione2007,Weiss2022}.

For weak system-bath interactions characterized by a small parameter $\lambda$, bath-induced dissipation and fluctuation effects emerge at the order of $\lambda^2$. 
Applying the Born-Markov approximation at this level yields the Redfield equation~\cite{Redfield1957}, 
which generally violates complete positivity~\cite{TupkaryEtAl2022} and detailed balance~\cite{SoretCavinaEsposito2022}. 
The subsequent secular (rotating-wave) approximation leads to the Davies equation by discarding off-diagonal interferences~\cite{Davies1974,Alicki1976}, 
and this recovers the complete positivity and a Gelfand--Naimark--Segal (GNS) detailed balance. 
However, that secular approximation typically fails in open many-body systems, 
due to the dense energy spectra~\cite{MozgunovLidar2020,Trushechkin2021}.
This problem motivates extensive works
on nonsecular quantum master equations~\cite{SchallerBrandes2008,PtaszynskiEsposito2019,NathanRudner2020,Davidovic2020,BeckerWuEckardt2021,WinczewskiEtAl2024,ShiraishiEtAl2025,ShiraishiNakagawaMori2026,Schnell2025,TabaneraBravoEtAl2026,BoubakourEtAl2026,CattaneoEtAl2019,FarinaGiovannetti2019,Trushechkin2021,PottsKalaeeWacker2021}, but existing schemes
fail to simultaneously achieve two key requirements: (i) preserving the detailed balance structure (ii) maintaining time-uniform $\mathcal{O}(\lambda^2)$ accuracy.

The detailed-balance requirement reflects the microscopic reversibility of the system and its thermal environment~\cite{Agarwal1973,CarmichaelWalls1976,Alicki1976,EspositoHarbolaMukamel2009,SoretCavinaEsposito2022}.    At thermal equilibrium, there are time-reversal relations on the fluctuations and response of system observables, 
which an effective description should preserve after eliminating the bath~\cite{Onsager1931,Kubo1957,Tsang2024}. 
In classical stochastic dynamics, it is expressed by detailed balance: in equilibrium state, 
a trajectory and its time-reversed counterpart occur with equal probability. 
In quantum regime, Kubo--Martin--Schwinger (KMS) detailed balance generalizes GNS detailed balance and accommodates nonsecular dynamics. It can be formulated with or without explicit time reversal, 
giving the time-reversal KMS (TR-KMS) and non-time-reversal KMS (nTR-KMS) conditions, respectively~\cite{FagnolaUmanita2007,FagnolaUmanita2011}. 
KMS detailed balance has recently attracted significant attention because of its role in quantum Gibbs sampling~\cite{ChenKastoryanoGilyen2023,DingLiLin2025,ChenEtAl2025,GilyenEtAl2026,ChenDingZhang2026,OngEtAl2026} 
and quantum thermodynamics~\cite{Kwon2019,Aaberg2018,Chen2025PRB,Chen2026PRE}.
The microscopic origin and physical interpretation of KMS detailed
balance for nonsecular many-body dynamics have recently begun to
be explored~\cite{ScandiAlhambra2026}.

The second requirement is $O(\lambda^2)$ accuracy in the reduced
density matrix relative to the microscopic evolution, concerning the reliability of
the master equation. 
In the weak-coupling setting considered here, the Redfield equation achieves
this accuracy on the dissipative timescale $\tau_d\sim\lambda^{-2}$
although an estimate on this timescale alone does not ensure that the error remains controlled at long times.
In the regime where the secular approximation is valid, time-uniform bounds are
known for Davies dynamics~\cite{Merkli2022}, including an $O(\lambda^2)$
bound for sufficiently rapidly decaying bath correlations~\cite{Merkli2020}.
Beyond secular regime, time-uniform error bounds have also been
explored recently for several nonsecular master equations~\cite{IkeuchiMori2025,ikeuchi2026}.
Nevertheless, to our knowledge, a time-uniform $O(\lambda^2)$ error bound 
relative to the microscopic dynamics has yet to be established
for a completely positive master equation
when the secular approximation fails.

\begin{table*}[!t]
\centering
\caption{
Comparison of selected thermal quantum master equations.
Here $\Delta\rho(t)$ denotes the trace-norm error relative
to the microscopic reduced dynamics, and
$\tau_d\sim\lambda^{-2}$ is the relaxation timescale.
The estimates hold under the respective assumptions of each analysis.
}
\label{tab:comparison}
\renewcommand{\arraystretch}{1.5}
\begin{tabular*}{\textwidth}{
    @{\extracolsep{\fill}} l c c c @{}
}
\hline\hline
\noalign{\vskip2pt}
Dynamics
& \shortstack{Nonsecular Interference}
& \shortstack{Detailed-balance\\structure}
& \shortstack{Microscopic\\state-error estimate}
\\[2pt]
\hline
\noalign{\vskip2pt}
Davies
& Discarded
& \shortstack{GNS\textsuperscript{a}\\(TR-KMS and nTR-KMS)}
& \shortstack[l]{
  $\displaystyle \sup_{t\geq0}\Delta\rho(t)=O(\lambda^2)$ (secular)~\cite{Merkli2020}\\
  $\displaystyle \sup_{t\geq0}\Delta\rho(t)=O(1)$ (nonsecular)}
\\[2pt]
\hline
\noalign{\vskip2pt}
Scandi--Alhambra (SA)~\cite{ScandiAlhambra2026}
& Retained
& nTR-KMS
& $\displaystyle \Delta\rho(t)
   =O\!\left[\lambda\left(1+\frac{t}{\tau_d}\right)\right]$
\\[2pt]
\hline
\noalign{\vskip2pt}
\textbf{TR-KMS completion (this work)}
& \textbf{Retained}
& \textbf{TR-KMS}
& $\displaystyle \boldsymbol{
   \sup_{t\geq0}\Delta\rho(t)=O(\lambda^2)}$
\\[2pt]
\hline\hline
\end{tabular*}

\par\smallskip
\begin{minipage}{\textwidth}
\footnotesize\raggedright
\textsuperscript{a} GNS detailed balance implies the nTR-KMS condition for the Davies generator. If the microscopic composite dynamics is invariant under time reversal, GNS detailed balance also implies TR-KMS detailed balance for the Davies generator.
\end{minipage}

\end{table*}

An advance toward reconciling these two requirements
was recently made in Ref.~\cite{ScandiAlhambra2026}, where 
Scandi and Alhambra apply temporal
coarse-graining of the reduced density matrix to obtain a nonsecular
master equation satisfying nTR-KMS detailed balance. We refer to it as
the Scandi--Alhambra (SA) equation.
However, its reported error bound scales as $\Delta\rho(t)\lesssim C \lambda(1+t/\tau_d)$, 
where $\tau_d$ is the relaxation timescale. 
This falls short of the time-uniform $O(\lambda^2)$ accuracy required here.
Even on the relaxation timescale, the bound is $O(\lambda)$, 
weaker than the $O(\lambda^2)$ accuracy of second-order Redfield dynamics. 
Two further limitations concern its equilibrium state 
and the freedom in its construction: 
the stationary Gibbs state of the renormalized Hamiltonian 
does not generally reproduce the second-order mean-force correction to the reduced equilibrium state, 
and the adjustable 
coarse-graining window leads to a family of effective generators.

In this paper, we construct a completely positive, nonsecular quantum master equation that 
fulfills both requirements: 
TR-KMS detailed balance and a time-uniform $O(\lambda^2)$ error bound relative to the microscopic dynamics. 
The resulting evolution, which we call \emph{TR-KMS completion dynamics}, 
describes relaxation beyond the regime of validity of the secular approximation.
Moreover, its stationary state agrees with the exact mean-force Gibbs state up to $O(\lambda^4)$, 
and for an interaction decomposition $V=\sum_{i=1}^n A_i\otimes B_i$, 
its dissipative part is represented by at most $n$ Lindblad operators. 
This equation is almost uniquely fixed within our construction, 
with a canonical choice resolving the remaining freedom in the multichannel case.

The key idea in the construction is to perform a detailed-balance completion of the Redfield generator. 
We choose the thermal reference from the second-order Hamiltonian of mean force~\cite{TrushechkinMerkliCresserAnders2022,CresserAnders2021,TimofeevTrushechkin2022} and construct 
coefficients that satisfy TR-KMS while matching the Redfield values at resonance. 
In our construction, the mean-force Hamiltonian not only captures the leading correction to thermal equilibrium but also plays a dynamical role, making resonance matching compatible with TR-KMS detailed balance.
For a fixed finite-dimensional system, the error bound analysis is based on two physical conditions:
controlled microscopic Redfield error~\cite{ScandiAlhambra2026} and fast relaxation of the TR-KMS completion dynamics.

Our construction highlights a dynamical distinction between TR-KMS and nTR-KMS detailed balance condition. Within our completion procedures of the Redfield equation, TR-KMS allows detailed balance to be restored while preserving second-order dynamical accuracy, whereas nTR-KMS is generically incompatible with the conditions required to retain this accuracy. 
Our completion achieves a time-uniform \(O(\lambda^2)\) state-error bound relative to microscopic dynamics. For comparison, the error bound established for the SA equation~\cite{ScandiAlhambra2026} scales only as \(O(\lambda)\) on the relaxation timescale \(t\sim\tau_d\).
These results suggest that TR-KMS could be better suited than nTR-KMS to constructing master equations that faithfully reproduce microscopic nonsecular dynamics.

Table~\ref{tab:comparison} compares TR-KMS completion with
established quantum master equations satisfying detailed balance.
Among these approaches, only TR-KMS completion guarantees a time-uniform
$O(\lambda^2)$ state-error bound relative to the microscopic dynamics
while retaining nonsecular interference between distinct transitions.

We further test TR-KMS completion against exact
microscopic results for two coupled harmonic oscillators interacting
with thermal reservoirs (Sec.~\ref{sec:oscillator-benchmark}). By
varying their frequency splitting, we compare secular, intermediate,
and nonsecular regimes. TR-KMS completion retains the $O(\lambda^2)$
dynamical accuracy. In comparison, the 
SA and Davies equations can develop $O(\lambda)$ and $O(1)$ dynamical errors, respectively, in different frequency regimes.
On the relaxation timescale, this benchmark reproduces the error
scalings summarized in Table~\ref{tab:comparison}.

The resulting framework provides a thermodynamically consistent description of how quantum interference shapes relaxation in open many-body systems, 
with dynamical accuracy controlled throughout the relaxation process. 
Its explicitly constructed generator offers a starting point for studying thermodynamics in systems 
whose dense transition spectra preclude a fully secular description.

\subsection*{Outline of the paper}

\begin{figure}
 \centering
 \includegraphics[width=\columnwidth]{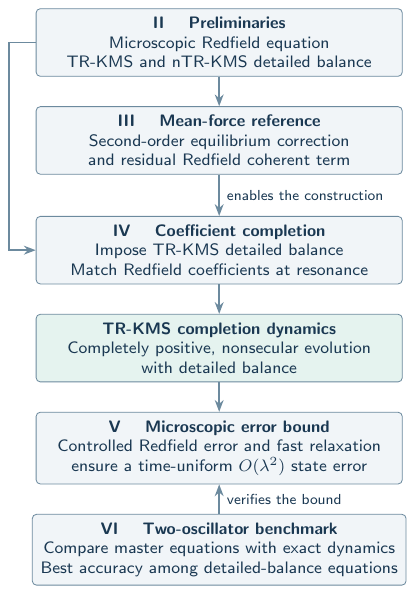}
 \caption{Roadmap of the paper.
}
 \label{fig:construction-roadmap}
\end{figure}

Figure~\ref{fig:construction-roadmap} shows how we construct the
TR-KMS completion, establish its error bound, and benchmark it
against the exact dynamics in an example.

Section~\ref{sec:preliminary} reviews the microscopic derivation of the Redfield equation and expresses its dissipative and coherent coefficients in the Bohr-frequency representation. It then introduces the TR-KMS and nTR-KMS detailed-balance conditions and gives their corresponding constraints on the generator coefficients. 
We also show that the expression for the coherent correction in TR-KMS and nTR-KMS can be unified.

Our construction then proceeds in two steps.

First, in Sec.~\ref{sec:reference},
we choose the reference Hamiltonian from the second-order expansion of the Hamiltonian of mean force,
$H^{\mathrm{ref}}=H_0+\lambda^2H_2^{\mathrm{MF}}$.
This incorporates the leading interaction correction to thermal equilibrium. 
The resulting stationary state, \(\sigma=e^{-\beta H^{\mathrm{ref}}}/Z\), 
agrees with the mean-force Gibbs state through second order. 
With this mean-force reference, the Redfield equation already satisfies part of the KMS detailed-balance condition. This structure provides the basis for resonance matching in the next step.

Second, in Sec.~\ref{sec:completion},
we construct a positive dissipation kernel in factorized form,
\(K^K(\omega',\omega)=Q(\omega')^\dagger Q(\omega)\), 
and impose detailed balance to obtain the coherent coefficient $h^K(\omega^\prime,\omega)$, 
which should be continuous.
We further require the dissipative and coherent coefficients to match their Redfield counterparts at resonance, $K^K(\omega,\omega)=K^R(\omega,\omega)$ and $h^K(\omega,\omega)=h^R(\omega,\omega)$.
Here $K^R$ and $h^R$ are Redfield coefficients after adopting reference Hamiltonian $H^{\rm ref}$.
Within this construction, TR-KMS permits a solution of the above continuous resonance matching, whereas nTR-KMS is generically incompatible with these matching conditions.

The resulting TR-KMS completion equation has the Gorini-Kossakowski-Sudarshan-Lindblad (GKSL) form, 
and it satisfies TR-KMS detailed balance with respect to \(\sigma\). 
No secular approximation is used, and
interference between different transition frequencies is retained.

Section~\ref{sec:error} analyzes the accuracy relative to
microscopic dynamics for a fixed finite-dimensional system. The error bound rests on two physical conditions:
controlled microscopic Redfield error 
and rapid mixing condition.
The first assumption has been proved for Gaussian bath with fast-decaying time correlation \cite{ScandiAlhambra2026,AgerskovNathan2026}. 
The second condition holds for dynamics that relaxes on the dissipative timescale.
Under these conditions, solutions starting from the same initial
density matrix satisfy
\begin{equation}
\sup_{t\ge0}
\|\rho_{\mathrm{mic}}(t)-\rho_K(t)\|_1
\le C\lambda^2 .
\end{equation}
with $C$ independent of time and sufficiently small $\lambda$.

In Sec.~\ref{sec:oscillator-benchmark}, we benchmark the master equations
against the exact microscopic dynamics of two coupled oscillators
interacting with thermal reservoirs. We denote the frequency separation of the two eigenmode as $\Delta$. The model allows us to test accuracy in secular ($\Delta\gg \lambda^2$), intermediate ($\Delta\propto \lambda$), and nonsecular ($\Delta\propto \lambda^2$) regimes. 
TR-KMS completion retains $O(\lambda^2)$ dynamical accuracy across
all three scans relative to the exact reduced equilibrium state. 
It achieves the most favorable
error scaling among the existing detailed-balance master equations tested there.

\section{Preliminary}\label{sec:preliminary}

We first review the microscopic derivation and coefficient
conventions of the Redfield equation, then introduce the general
definitions of quantum detailed balance.

\subsection{Derivation of Redfield equation}\label{sec:prelim-redfield}
The starting point is a system weakly coupled to a thermal bath,
characterized by the Hamiltonian of the composite system, 
\begin{equation}
    \tilde H=H_0+H_B+\lambda V,
    \qquad V=\sum_i A_i\otimes B_i,
\end{equation}
where $H_0$ and $H_B$ are the Hamiltonians of the uncoupled system and
bath, respectively, and $V$ is their interaction. The dimensionless
parameter $\lambda$ controls the coupling strength, and the channel
operators $A_i$ and $B_i$ are Hermitian system and bath operators,
respectively. We denote the total density
operator by $\tilde\rho$ and the reduced state by
$\rho=\Tr_B\tilde\rho$. 
The uncoupled bath equilibrium state is
$\rho_0^B=\ue^{-\beta H_B}/Z_B$, with inverse temperature $\beta$.
We center the bath operators, $\Tr_B(\rho_0^B B_i)=0$, by absorbing any
nonzero mean into $H_0$.

We denote the antiunitary time-reversal operator acting on state vectors
by $\Theta$. The corresponding superoperator acts on density operators as
\begin{equation}\label{eq:db-time-reversal}
    \mathcal T(\rho)=\Theta \rho\Theta^{-1}.
\end{equation}
We assume that time reversal acts separately on the system and bath
and that the microscopic Hamiltonian satisfies the time-reversal invariance, 
$\mathcal T\tilde H=\tilde H$.
The invariance of $V$ allows us to choose a Hermitian channel
decomposition in which the system and bath channels transform with
the same real orthogonal matrix $E$:
\begin{equation}\label{in-TR}
    \mathcal T A_i=\sum_j E_{ij}A_j,
    \quad \mathcal T B_i=\sum_j E_{ij}B_j,
    \quad E^TE=I.
\end{equation}
An orthogonal change of
channel basis puts $E$ in the block-diagonal form
\begin{equation}\label{E_s}
    E=\operatorname{diag}(1,\ldots,1,-1,\ldots,-1,J,\ldots,J),
\end{equation}
with two-dimensional matrix 
\begin{equation}
    J=\begin{pmatrix}0&1\\-1&0\end{pmatrix}.
\end{equation}
For the usual time-reversal conventions, 
$\Theta^2=(-1)^F$, where $F$ is the fermion
number. Bosonic channels and even fermionic channels obey
$\mathcal T^2(A)=A$,
corresponding to the diagonal entries $+1$ or $-1$. In contrast, odd
fermionic channels obey $\mathcal T^2(A)=-A$ and form pairs represented
by the $J$ blocks.

The microscopic evolution equation for the composite system is given by
\begin{equation}
 \begin{aligned}
    \dot{\tilde\rho}&=\tilde{\mathcal L}\tilde\rho,
    \qquad \tilde{\mathcal L}=\tilde{\mathcal L}_0
                                  +\lambda\tilde{\mathcal L}_1,
 \end{aligned}
\end{equation}
with the generator separated into 
\begin{equation}
    \tilde{\mathcal L}_0=-\ui\operatorname{ad}_{H_0+H_B},
    \qquad \tilde{\mathcal L}_1=-\ui\operatorname{ad}_V.
\end{equation}
Here $\operatorname{ad}_X(Y)=[X,Y]=XY-YX$ denotes the commutator
superoperator generated by $X$.
The coupling term $\tilde{\mathcal L}_1$ will be treated perturbatively
in $\lambda$.
We have used the generator with tilde to denote the composite system-bath dynamics 
and will use the generator without tilde to represent the dynamics for system only.

To eliminate the bath, we define the following projection operator 
\begin{equation}
    \mathcal P\tilde\rho=(\Tr_B\tilde\rho)\otimes\rho_0^B,
    \qquad \mathcal Q=1-\mathcal P.
\end{equation}
The projection satisfies $\mathcal P^2=\mathcal P$,
$[\mathcal P,\tilde{\mathcal L}_0]=0$, and
$\mathcal P\tilde{\mathcal L}_1\mathcal P=0$. 
The exact projected equation is~\cite{BreuerPetruccione2007}
\begin{equation}
 \begin{aligned}
    \pdv{t}\mathcal P\tilde\rho(t)
    &=\mathcal P\tilde{\mathcal L}\mathcal P\tilde\rho(t)
      +\mathcal P\tilde{\mathcal L}
       \ue^{t\mathcal Q\tilde{\mathcal L}}\mathcal Q\tilde\rho(0)\\
    &\quad+\int_0^t\ud s\,\mathcal P\tilde{\mathcal L}
      \ue^{s\mathcal Q\tilde{\mathcal L}}
      \mathcal Q\tilde{\mathcal L}\mathcal P\tilde\rho(t-s).
 \end{aligned}
\end{equation}
There are three terms on the right-hand side.
After tracing out the bath, the first term gives
$-\ui[H_0,\rho(t)]$. 
The second initial-correlation term vanishes for an initial product state
$\tilde\rho(0)=\rho(0)\otimes\rho_0^B$; 
for ordinary correlated state, 
it typically decays on the bath relaxation timescale 
and is negligible.

We expand the third term to second order in $\lambda$.
Assuming the bath memory time $\tau_B$ is much shorter than the system
relaxation time $\tau_d$, we retain only free system evolution during
the memory interval. For $t\gg\tau_B$, extending the integral to
infinity gives the Born--Markov approximation~\cite{BreuerPetruccione2007}
\begin{equation}\label{3rdterm}
 \begin{aligned}
    &\Tr_B\int_0^t\ud s\,\mathcal P\tilde{\mathcal L}
      \ue^{s\mathcal Q\tilde{\mathcal L}}
      \mathcal Q\tilde{\mathcal L}\mathcal P\tilde\rho(t-s)\approx \\
      & -\lambda^2\int_0^\infty\ud s\,
       \Tr_B[V,[V(-s),\rho(t)\otimes\rho_0^B]],
 \end{aligned}
\end{equation}
where $V(-s)=\ue^{-\ui(H_0+H_B)s}V\ue^{\ui(H_0+H_B)s}$ is in the interaction picture.
We assume that odd bath coupling correlations vanish, so the leading
omitted interaction order is $O(\lambda^4)$. 
The subsequent steps algebraically rewrite this second-order bath
contribution in terms of Bohr operators and bath spectra.

For the coefficient conventions, decompose each coupling operator $A_i$ into
Bohr operators of the bare Hamiltonian,
\begin{equation}\label{eq:prelim-bare-Bohr}
 \begin{aligned}
    A_i&=\sum_{\omega\in\Omega_0} A_i^0(\omega),
    & [H_0,A_i^0(\omega)]&=-\omega A_i^0(\omega),\\
    A_i^0(\omega)^\dagger&=A_i^0(-\omega),
    & A_i^0(\omega;-s)&=\ue^{\ui\omega s}A_i^0(\omega).
 \end{aligned}
\end{equation}
Here $A^0(\omega)$ can be regarded as the eigenvectors
of the commutation operation $\operatorname{ad}_{H_0}$, and
$\omega$ runs over the set of eigen energy differences
$\Omega_0=\{E_n^0-E_m^0\}$ of $H_0$.

Tracing out the bath leaves the bath correlation functions,
\begin{equation}
    C_{ij}(s)=\ev{B_i(s)B_j(0)}_0
       =\Tr_B[\rho_0^B B_i(s)B_j(0)],
\end{equation}
where the average is taken in the uncoupled bath dynamics.
Its one-sided Fourier transform reads,
\begin{equation}\label{eq:prelim-bath-transform}
    \int_0^\infty\ud s\,\ue^{\ui\omega s}C_{ij}(s)
       =\frac12\gamma_{ij}(\omega)+\ui S_{ij}(\omega),
\end{equation}
where we decompose it to two Hermitian matrices $\gamma(\omega)$ and $S(\omega)$.
We consider the usual case in which $\gamma(\omega)$ is positive definite and hence invertible at the relevant frequencies.
Time-reversal symmetry and thermal equilibrium~\cite{Kubo1957,MartinSchwinger1959} of the bath respectively imply
\begin{equation}\label{gamma_r}
 \begin{aligned}
    &S(\omega)=ES(\omega)^TE^T,\quad
    \gamma(\omega)=E\gamma(\omega)^T E^T,\\
    &\gamma(\omega)=\ue^{\beta\omega}\gamma(-\omega)^T.
 \end{aligned}
\end{equation}
These bath relations encode the microscopic thermal reversibility.

Using these transforms and rearranging the terms gives the Redfield
equation~\cite{Redfield1957},
\begin{equation}\label{eq:prelim-Redfield}
    \dot\rho=-\ui[H_0,\rho]+\lambda^2\mathcal L^R_2[\rho],
\end{equation}
where the second-order generator induced by the heat bath is
\begin{equation}\label{eq:Redfield-2ndgenerator}
\begin{aligned}
    \mathcal L^R_2[\rho]=& -\ui\sum_{ij;\omega'\omega\in\Omega_0}
        r^\text{LS}_{ij}(\omega',\omega)\comm{P^0_{ij;\omega'\omega}}{\rho}\\
        &+   \sum_{ij;\omega'\omega\in\Omega_0}
        K^R_{ij}(\omega',\omega)
        \mathcal D^0_{ij;\omega'\omega}[\rho].
        \end{aligned}
\end{equation}
We have used two shorthand notations: an operator product $P^0_{ij;\omega'\omega}$ and a dissipative
superoperator $\mathcal D^0_{ij;\omega'\omega}$,
\begin{equation}\label{eq:prelim-bare-PD}
 \begin{aligned}
    P^0_{ij;\omega'\omega}
       &=A_i^0(\omega')^\dagger A_j^0(\omega),\\
    \mathcal D^0_{ij;\omega'\omega}[\rho]
       &=A_j^0(\omega)\rho A_i^0(\omega')^\dagger
          -\frac12\pb{P^0_{ij;\omega'\omega}}{\rho}.
 \end{aligned}
\end{equation}
In Eq.~\eqref{eq:Redfield-2ndgenerator}, 
the first term represents a coherent evolution under the Lamb shift Hamiltonian
\begin{equation}\label{lamb-shift}
    H^\text{LS} = \sum_{ij;\omega'\omega\in\Omega_0}
        r^\text{LS}_{ij} P^0_{ij;\omega'\omega},
\end{equation}
while the second term represents the dissipative evolution.
The Lamb-shift coefficient and dissipative coefficient matrices are
\begin{equation}\label{eq:prelim-Redfield-coefficients}
 \begin{aligned}
    K^R(\omega',\omega)
       &=\frac12[\gamma(\omega)+\gamma(\omega')]
                       +\ui[S(\omega)-S(\omega')],\\
    r^\text{LS}(\omega',\omega)
       &=\frac12[S(\omega)+S(\omega')]
                       -\frac\ui4[\gamma(\omega)-\gamma(\omega')].
 \end{aligned}
\end{equation}
The Redfield equation retains terms with $\omega\ne\omega'$ in both
the dissipative and Lamb-shift contributions. These terms
describe interference between transitions with different Bohr frequencies.

The Redfield equation has two structural shortcomings. First, the
kernel $K^R$ is generally not positive semidefinite, so the evolution
need not be completely positive and can produce density matrices with
negative eigenvalues~\cite{Davidovic2020,HartmannStrunz2020}. This also prevents the direct use of the conventional quantum-jump unraveling~\cite{WeimerEtAl2021,BeckerNetzerEckardt2023}.
Second, the dynamics generally fails to satisfy the quantum detailed
balance. Detailed balance expresses thermal reversibility beyond a Gibbs stationary
state at the level of effective dynamics. Recovering this structure is 
essential to a thermodynamically consistent description of
relaxation~\cite{Alicki1976,RobertsLingenfelterClerk2021}.

\subsection{Secular approximation and its scope}\label{subsec:Secular approximation}
For systems with well separated transition frequencies $\omega$, 
the secular approximation (or rotating wave approxmation) resolves both problems.
In Eq.~\eqref{eq:prelim-bare-PD}, non-resonant terms (with $\omega\ne\omega'$) oscillate
at frequency $\abs{\omega-\omega^\prime}$. 
For the relevant distinct Bohr frequencies, they can be averaged out 
if the dissipation timescale $\tau_d$ (which scales as $\lambda^{-2}$) satisfies
\begin{equation}\label{eq:prelim-secular-window}
    \abs{\omega-\omega'}\tau_d\gg1.
\end{equation}
Only terms with the same frequency are retained after the secular approximation.
The resulting Davies equation is given by the following coefficients, 
\begin{equation}
 \begin{aligned}
    K^D(\omega',\omega)&=\gamma(\omega)\delta_{\omega'\omega},\\
    r^D(\omega',\omega)&=S(\omega)\delta_{\omega'\omega}.
 \end{aligned}
\end{equation}
Under the secular approximation, only coherences between degenerate
energy eigenstates can couple to populations~\cite{BulnesCuetaraEspositoSchaller2016}.
For the Davies equation, positivity of $\gamma(\omega)$
guarantees complete positivity. 
The relation $\gamma(\omega)=\gamma(\omega)\ue^{\beta\omega}$ in
Eq.~\eqref{gamma_r}, together with the vanishing of non-resonant terms, ensures that the dynamics satisfies Gelfand--Naimark--Segal (GNS) detailed balance with respect to the bare Gibbs state
$\ue^{-\beta H_0}/\Tr\ue^{-\beta H_0}$~\cite{Davies1974,Alicki1976,KossakowskiEtAl1977}.
GNS detailed balance is a special case of the KMS detailed balance that will be introduced below.

However, secular approximation is invalid for quantum many-body systems. 
The relevant Bohr frequencies
can become extremely dense, which scales exponentially with the system size. 
If $|\omega-\omega'|\lesssim\tau_d^{-1}$, interference between
nearby transitions persists throughout relaxation and can affect both
populations and coherences~\cite{Trushechkin2021,ScandiAlhambra2026}.
Moreover, exact frequency resolution can turn local couplings into
nonlocal jump operators, complicating many-body
simulation~\cite{ChenEtAl2025}.

To formulate a detailed-balance completion beyond the secular
approximation, we next introduce the general definitions of quantum
detailed balance and their constraints on an effective master equation.

\subsection{Quantum detailed balance}
In classical open systems described by stochastic dynamics, detailed
balance expresses time-reversal symmetry for trajectory probability, characterizing thermal equilibrium at the level of stochastic dynamics.

For quantum master equations, there are two forms of
Kubo--Martin--Schwinger (KMS) detailed balance used in
literature~\cite{FagnolaUmanita2007,FagnolaUmanita2011}.
Time-reversal KMS
(TR-KMS) retains the time-reversal structure of classical detailed
balance~\cite{DuvenhageEtAl2025,BenoistEtAl2026}, whereas non-time-reversal KMS (nTR-KMS) is defined without introducing explicit time
reversal. 
The connection between TR-KMS and classical detailed
balance is discussed in Appendix~\ref{app:db}. Below, we introduce both conditions and their
constraints on dissipative and coherent evolution.

We now discuss quantum detailed balance as a structural condition on
a general GKSL generator~\cite{GKS1976,Lindblad1976}. Consider a
time-independent Markov quantum master equation,
\begin{equation}
    \pdv{t} \rho  = \mathcal L \rho,
\end{equation}
where $\mathcal L$ is its generator.
We take its stationary state and write it in Gibbs form,
\begin{equation}
    \sigma=\ue^{-\beta H}/Z,\qquad Z=\Tr\ue^{-\beta H}.
\end{equation}
Here $H$ specifies the reference equilibrium state, rather than
necessarily the full coherent Hamiltonian in $\mathcal L$. 
With the time reversal defined in Eq.~\eqref{eq:db-time-reversal},
we consider the case of time-reversal invariant Hamiltonian, $\mathcal T H=H$.

We define the thermal
multiplication map
\begin{equation}\label{eq:db-Msigma}
    M_\sigma(X)=\sigma^{1/2}X\sigma^{1/2},
\end{equation}
and the adjoint generator $\mathcal L^\sharp$ from
\begin{equation}\label{eq:db-adjoint}
    \Tr[X^\dagger\mathcal L(Y)]
       =\Tr[(\mathcal L^\sharp(X))^\dagger Y],
\end{equation}
where $X$ and $Y$ are arbitrary operators with finite trace.
The trace-preserving property of the dynamics
implies $\mathcal L^\sharp(I)=0$.

The quantum master equation satisfies the TR-KMS detailed balance
if~\cite{FagnolaUmanita2011}
\begin{equation}\label{TR-KMS}
    \mathcal L=M_\sigma\mathcal T\mathcal L^\sharp
                       \mathcal T^{-1}M_\sigma^{-1}.
\end{equation}
This relation implies
$\mathcal L\sigma=0$ and imposes a strong strcuture on the
dynamics. 
In the current setup, $\mathcal T$ and $M_\sigma$ are interchangeable. 

This detailed balance condition becomes explicit in the
energy basis of the reference Hamiltonian.
We express the general generator $\mathcal L$ in a Hermitian basis
$\{A_i\}$, decomposed into Bohr operators $A_i(\omega)$ of the reference
Hamiltonian $H$.
The action of $\mathcal T$ is
represented by a real orthogonal matrix $E$, just as in Eq.~\eqref{in-TR}.
We use shorthand notations $P_{ij;\omega'\omega}$ and
$\mathcal D_{ij;\omega'\omega}$ for the products and dissipators defined
as in Eq.~\eqref{eq:prelim-bare-PD}, with $A_i(\omega)$ in place
of $A_i^0(\omega)$.

In this representation, the general GKSL generator takes the form
\begin{equation}\label{generator}
    \mathcal L\rho=-\ui[H+N+G,\rho]
      +\sum_{ij;\omega'\omega\in\Omega_0}K_{ij}(\omega',\omega)
                        \mathcal D_{ij;\omega'\omega}[\rho],
\end{equation}
where
$G$ denotes the coherent correction,
\begin{equation}
    G=\sum_{ij\omega'\omega}h_{ij}(\omega',\omega)
                                   P_{ij;\omega'\omega}.
\end{equation}
TR-KMS allows an arbitrary Hermitian operator $N$ satisfying $[H,N]=0$ and $\mathcal T N=N$. 
For other terms, TR-KMS imposes
\begin{align}
    K(\omega',\omega)
       &=\ue^{\beta(\omega+\omega')/2}
                       E K(-\omega',-\omega)E^T,
                         \label{TR-KMSe1}\\
    h(\omega',\omega)
       &=f[K](\omega',\omega)
                                      \quad(\omega\ne\omega'),
                         \label{TR-KMSe2}
\end{align}
with $\nu=\omega-\omega'$ and 
\begin{equation}\label{eq:hmf-f}
\begin{aligned}
    f[K](\omega',\omega)
&=
\frac{\mathrm{i}\,e^{\beta(\omega+\omega')/2}}
     {2\sinh(\beta\nu/2)}
K(-\omega,-\omega')^{T}
\\&-
\frac{\mathrm{i}}{2}
\coth\!\left(\frac{\beta\nu}{2}\right)
K(\omega',\omega).
\end{aligned}
\end{equation}
The first relation constrains the ratio between a transition pair, 
with the thermal weight $\ue^{\beta(\omega^\prime+\omega)/2}$.
The second fixes the coherent coefficients to match the dissipative part.
For a dissipative coefficient $K$ satisfying Eq.~\eqref{TR-KMSe1},
Eq.~\eqref{TR-KMSe2} can equivalently be written as
\begin{equation}\label{eq:db-TR-explicit}
 \begin{aligned}
 h(\omega',\omega)
 &=\frac\ui2\csch\!\left(\frac{\beta\nu}{2}\right)
                  E K(\omega',\omega)^*E^T\\
 &-\frac\ui2\coth\!\left(\frac{\beta\nu}{2}\right)K(\omega',\omega),
 \end{aligned}
\end{equation}
for $\nu\ne0$.
Here $K^*$ denotes elementwise complex conjugation.
Appendix~\ref{app:db-coefficients} proves the equivalence and derives
these coefficient conditions.

We note that for $\omega=\omega'$, there is an arbitrary freedom in $h(\omega,\omega)$, 
since $h(\omega,\omega)P_{ij;\omega\omega}$ is commutative with $H$ and can be absorbed into $N$.
The freedom in $N$ is allowed 
because the evolution generated by $H+N$ independently satisfies
TR-KMS: antiunitarity gives
\begin{equation}
    \mathcal T(-\ui\operatorname{ad}_{H+N})^\sharp
       \mathcal T^{-1}=-\ui\operatorname{ad}_{H+N},
\end{equation}
and this coherent term commutes with $M_\sigma$.

The alternative nTR-KMS condition omits explicit time
reversal~\cite{DingLiLin2025}. Its strict form is
$\mathcal L=M_\sigma\mathcal L^\sharp M_\sigma^{-1}$, obtained by omitting
time reversal in Eq.~\eqref{TR-KMS}.
However, this prohibits any independent coherent evolution commuting
with $\sigma$, even including $-\ui [H,\rho]$.
In open-system applications, nTR-KMS is commonly used in an extended
form that retains this independent coherent evolution~\cite{FagnolaUmanita2007}:
\begin{equation}
 \begin{aligned}
    \mathcal L&=-\ui\operatorname{ad}_{H+N}+\mathcal S,\qquad [H,N]=0,\\
    \mathcal S&=M_\sigma\mathcal S^\sharp M_\sigma^{-1}.
 \end{aligned}
\end{equation}
In the energy representation, the coefficient relations for nTR-KMS detailed balance are
\begin{align}\label{eq:db-ntr-coefficients1}
    K(\omega',\omega)
       &=\ue^{\beta(\omega+\omega')/2}K(-\omega',-\omega)^*,\\
    h(\omega',\omega)
       &=f[K](\omega',\omega)\quad(\omega\ne\omega').\label{eq:db-ntr-coefficients2}
\end{align}
Here, the same functional $f[K]$ is given by Eq.~\eqref{eq:hmf-f}.
For a Hermitian kernel $K$ satisfying the first relation~\eqref{eq:db-ntr-coefficients1}, 
the second relation~\eqref{eq:db-ntr-coefficients2} is equivalent to the following explicit form used in Ref.~\cite{ChenKastoryanoGilyen2023}, 
\begin{equation}\label{eq:db-ntr-explicit}
 h(\omega',\omega)=-\frac\ui2\tanh\!\left(\frac{\beta\nu}{4}\right)
                     K(\omega',\omega)\quad(\nu\ne0).
\end{equation}
Appendix~\ref{app:db-ntr-coefficients} derives both forms.

Therefore, we note that TR-KMS and nTR-KMS share the same relation that fixes the coherent correction in terms of $f[K]$, 
and their difference lies in the relation that constrains the ratio between a transition pair; see Eqs.~\eqref{TR-KMSe1} and \eqref{eq:db-ntr-coefficients1}. 
On the right-hand sides, TR-KMS contains the time reversal $E KE^T$, while nTR-KMS uses the complex conjugate $K^*$.
The TR-KMS and nTR-KMS detailed balance relations coincide only in the special case where $EKE^T=K^*$.

\section{Reference Hamiltonian}\label{sec:reference}
We use the microscopic setup and notation introduced in
Sec.~\ref{sec:prelim-redfield} for the Redfield equation.
In this section, we specify the reference Hamiltonian, 
whose Gibbs state will give the steady state of the dynamics.
This step provides the basis for constructing the TR-KMS completion
dynamics in Sec.~\ref{sec:completion}.

When the microscopic system-bath dynamics relaxes to thermal
equilibrium, the exact reduced steady state is the mean-force Gibbs
state, obtained by tracing out the bath from the interacting Gibbs
state~\cite{TrushechkinMerkliCresserAnders2022,SubasiEtAl2012}.
The Hamiltonian of mean force, conventionally used to describe
open-system equilibrium~\cite{TalknerHanggi2020}, is given by
\begin{equation}\label{eq:reference-exact-HMF}
 \ue^{-\beta H^{\rm MF}}
 =\frac{\Tr_B\ue^{-\beta(H_0+H_B+\lambda V)}}{Z_B},
 \qquad Z_B=\Tr_B\ue^{-\beta H_B},
\end{equation}
so that $\rho_{\rm MF}\propto\ue^{-\beta H^{\rm MF}}$. 
This state incorporates the equilibrium correction due to the system-bath interaction~\cite{KonenbergMerkli2017,KonenbergMerkli2019Correction}.

In the current weak-coupling setup, the Hamiltonian of mean force
takes the form
\begin{equation}
 H^{\rm MF}=H_0+\lambda^2H_2^{\rm MF}+O(\lambda^4).
\end{equation}
We therefore choose its first two orders as our reference Hamiltonian,
\begin{equation}\label{eq:reference-Hamiltonian}
 H^{\rm ref}=H_0+\lambda^2H_2^{\rm MF},\qquad
 \sigma=\frac{\ue^{-\beta H^{\rm ref}}}{Z},
\end{equation}
where $Z=\Tr\ue^{-\beta H^{\rm ref}}$. 
In our detailed balance construction, $\sigma$ will be
the steady state of the constructed dynamics, which
differs from the exact mean-force state only in the fourth order,
$\sigma -\rho_\text{MF} =O(\lambda^4)$. 
Moreover, $H_2^{\rm MF}$ can be explicitly computed from the bare
system operators and bath correlations, as shown below.

A notable fact is that, although the Redfield equation
describes the dynamics to second order in $\lambda$, its steady state
does not generally match the mean-force Gibbs state at that order.
In the weak-coupling expansion at fixed $H_0$, the
off-diagonal elements of the Redfield
steady state between distinct energies agree with those of the
mean-force Gibbs state through order $\lambda^2$, whereas the diagonal
populations generally differ at that
order~\cite{FlemingCummings2011,ThingnaWangHanggi2012,LobejkoEtAl2024,BeckerSchnellThingna2022,SartipiEtAl2026}.
An explanation is given in Appendix~\ref{app:hmf}.
Perturbative analyses of several completely positive Redfield approximations also find equilibrium corrections that differ from the mean-force state~\cite{LeeYeo2022}.

To obtain the second-order correction explicitly, 
a direct thermal expansion of the mean-force
Hamiltonian~\cite{TimofeevTrushechkin2022} yields
\begin{equation}\label{eq:hmf-H2}
 H_2^{\rm MF}
 =\sum_{ij\omega'\omega}g_{ij}(\omega',\omega)
 A_i^0(\omega')^\dagger A_j^0(\omega),
\end{equation}
with Bohr operators $A_i^0(\omega)$ of $H_0$ and coefficients 
\begin{align}\label{eq:hmf-g}
 g_{ij}(\omega^\prime,\omega)&=\frac{\ue^{\beta(\omega-\omega^\prime)}J_{ij}(\omega^\prime)-J_{ij}(\omega)}
                   {1-\ue^{\beta(\omega-\omega^\prime)}},\\
 J(\omega)&=-S(\omega)-\ue^{\beta \omega}S(-\omega)^T.\label{eq:hmf-J}
 \end{align}
The derivation can be found in Refs.~\cite{LobejkoEtAl2024,TimofeevTrushechkin2022}, and it is also included in Appendix~\ref{app:hmf}. 
The above expression uses the same coefficient $S$ from bath correlations, as
the Redfield equation. 
The equal-frequency value $g_{ij}(\omega,\omega)$ is given by the
continuous limit.

We now rewrite the Redfield equation \eqref{eq:prelim-Redfield} by splitting out
the reference Hamiltonian $H^{\rm ref}=H_0+\lambda^2H_2^{\rm MF}$ from the total Lamb-shifted Hamiltonian $H_0+\lambda^2 H^\text{LS}$ in Eqs.~\eqref{eq:prelim-Redfield} and \eqref{lamb-shift}.
The following equation is equivalent to the original Redfield generator, 
\begin{equation}\label{eq:reference-Redfield}
\begin{aligned}
\mathcal{L}^{R}\rho={}&-\ui\comm{H^\text{ref}}{\rho}
\\&-\ui\lambda^{2}\sum_{ij;\omega'\omega\in\Omega_0}h_{ij}^{R}(\omega',\omega)\comm{P_{ij;\omega'\omega}^{0}}{\rho}\\&+\lambda^{2}\sum_{ij;\omega'\omega\in\Omega_0}K_{ij}^{R}(\omega',\omega)\mathcal{D}_{ij;\omega'\omega}^{0}[\rho].
\end{aligned}
\end{equation}
The residual coherent coefficient is obtained by subtracting the mean-force correction from the Redfield Lamb shift, which can be calculated explicitly,
\begin{equation}\label{eq:reference-hR}
 \begin{aligned}
 h^R(\omega',\omega)&=r^\text{LS}(\omega',\omega)-g(\omega',\omega)\\
                    &=f[K^R](\omega',\omega),\qquad \omega'\ne\omega.
 \end{aligned}
\end{equation}
Remarkably, we find that $h^R$ coincides with the compensation functional $f[K]$ defined in Eq.~\eqref{eq:hmf-f}.
See Eq.~\eqref{eq:hmf-f-eval} in Appendix~\ref{app:hmf-compatibility}
for the derivation.

The equal-frequency value $h^R(\omega,\omega)$ is given by the continuous limit of $\omega^\prime\rightarrow\omega$, which is finite.
To evaluate this continuous limit, we fix $\omega$ and set $\omega'=\omega-\nu$.
Expanding the Redfield kernel gives
\begin{equation}
 \begin{aligned}
 K^R(\omega-\nu,\omega)
 &=\gamma(\omega)+\nu\bigl[-\tfrac12\gamma'(\omega)+\ui S'(\omega)\bigr]
    +O(\nu^2).
 \end{aligned}
\end{equation}
For $K=K^R$ in Eq.~\eqref{eq:hmf-f}, the thermal
relation $\gamma(\omega)=\ue^{\beta\omega}\gamma(-\omega)^T$
cancels the constant terms. With the denominator written as
$\ui(\ue^{\beta\nu}-1)=\ui\beta\nu+O(\nu^2)$, the numerator is
$\ui\nu[S'(\omega)-\ue^{\beta\omega}S'(-\omega)^T]+O(\nu^2)$.
The equal-frequency limit is then
\begin{equation}\label{eq:reference-hR-resonance}
 h^R(\omega,\omega)=\lim_{\nu\to0}h^R(\omega-\nu,\omega)=\frac{S'(\omega)-\ue^{\beta \omega}S'(-\omega)^T}{\beta},
\end{equation}
where $S'(-\omega)$ denotes the derivative evaluated at $-\omega$.

From now on, we write $H\equiv H^{\rm ref}$ and use
its Bohr decomposition,
\begin{equation}
 A_i=\sum_{\omega\in\Omega} A_i(\omega),\qquad
 \comm{H}{A_i(\omega)}=-\omega A_i(\omega).
\end{equation}
Here, $\omega$ takes values from the set of eigen energy differences 
of the reference Hamiltonian, $\Omega=\{E_n-E_m\}$, which is usually different from $\Omega_0$ of the bare Hamiltonian. 
The superscript $0$ will continue to distinguish the bare Bohr operators $A^0_i(\omega)$ of $H_0$.

Now,
we make an approximation on the Redfield generator, by
replacing the bare Bohr operator $A_i^0(\omega)$ with the reference operator $A_i(\omega)$, 
\begin{equation}\label{eq:modified-Redfield}
 \begin{aligned}
 \mathcal{L}^{R}[\rho]=&-\ui[H,\rho]-\ui\lambda^{2}\sum_{ij;\omega'\omega\in\Omega}h_{ij}^{R}(\omega',\omega)[P_{ij;\omega'\omega},\rho]\\&+\lambda^{2}\sum_{ij;\omega'\omega\in\Omega}K_{ij}^{R}(\omega',\omega)\mathcal{D}_{ij;\omega'\omega}[\rho]+O(\lambda^4).
 \end{aligned}
\end{equation}
The product $P$ and dissipator $\mathcal D$ are
\begin{equation}\label{eq:completion-PD}
 \begin{aligned}
  P_{ij;\omega'\omega}
   &=A_i(\omega')^\dagger A_j(\omega),\\
  \mathcal D_{ij;\omega'\omega}[\rho]
   &=A_j(\omega)\rho A_i(\omega')^\dagger-\frac12\pb{P_{ij;\omega'\omega}}{\rho}.
 \end{aligned}
\end{equation}
They contain $A_i(\omega)$ with $\omega\in\Omega$ and should be distinguished with $P^0$ and $\mathcal D^0$ that use the bare Bohr operators $A_i^0(\omega)$ with $\omega\in\Omega_0$.
This replacement
generates an error with $O(\lambda^4)$ in the generator, which is negligible in a second-order description. 
The reason is that for any smooth function $g(\omega)$, 
two combinations $\sum_{\omega\in\Omega} g(\omega)A_i(\omega)$ and $\sum_{\omega\in\Omega^\prime} g(\omega) A_i^0(\omega)$ 
are close to each other, with difference given by $O(\lambda^2)$, 
while the coherent correction and the dissipation already include $\lambda^2$ in their strengths.

Choosing the second-order Hamiltonian of mean force $H=H+H_2^\text{MF}$ as the reference does more than providing a good steady state.
After using the reference Hamiltonian and the reference Bohr operators, the Redfield equation \eqref{eq:modified-Redfield} already obeys the second relation in TR-KMS (as well as nTR-KMS), 
\begin{equation}\label{eq:reference-shared-compensation}
 h^R=f[K^R],\qquad \omega'\ne\omega,
\end{equation}
according to Eq.~\eqref{eq:reference-hR}.
Therefore, the Redfield equation automatically satisfies ``half of'' the TR-KMS detailed balance, and this structure is apparent only when choosing the Hamiltonian of mean force as the reference. 
On the other hand, the Redfield kernel $K^R$ generally violates the first
relation~\eqref{TR-KMSe1} in TR-KMS detialed balance. In the following, we give a systematic modification of these coefficients
to satisfy the whole detailed balance.

\section{completion of the coefficients}\label{sec:completion}
We aim to modify the coefficients in Redfield equation to satisfy the full quantum detailed balance.
We seek a new TR-KMS completion generator $\mathcal{L}^{K}$, expressed in the energy basis,
\begin{equation}\label{eq:completion-ansatz}
 \begin{aligned}
 \mathcal{L}^{K}[\rho]=&-\ui[H^{{\rm ref}},\rho]-\ui\lambda^{2}\sum_{ij;\omega'\omega\in\Omega}h_{ij}^{K}(\omega',\omega)[P_{ij;\omega'\omega},\rho]\\&+\lambda^{2}\sum_{ij;\omega'\omega\in\Omega}K_{ij}^{K}(\omega',\omega)\mathcal{D}_{ij;\omega'\omega}[\rho].
 \end{aligned}
\end{equation}
Coefficients $K^K$ and $h^K$ uniquely specify the dynamics.
We seek coefficients $K^K$ and $h^K$ that satisfy detailed balance, 
while keeping the constructed dynamics sufficiently close to Redfield
to preserve its accuracy relative to the microscopic
dynamics.

To make the modified equation close to Redfield, 
we start from the following observation.
The secular approximation keeps only equal-frequency terms with $\omega^\prime=\omega$, as
interference between well-separated transitions averages out over the
relaxation time. For systems with dense spectra, however, interference
between near-resonant transitions must also be retained. These resonant and near-resonant contributions dominate the accumulated
bath-induced dynamics. This motivates matching the coefficients at resonance.

We impose the \emph{resonance
matching}: the generator $\mathcal L^K$ should match Redfield at resonance $\omega^\prime=\omega$,
and this match should be continuous when $\omega^\prime\to\omega$.
This requires both coefficients agree with their Redfield counterparts
at equal frequencies,
\begin{equation}\label{resonance_matching}
    K^K(\omega,\omega)=K^R(\omega,\omega),\quad
    h^K(\omega,\omega)=h^R(\omega,\omega),
\end{equation}
and these functions should be continuous at $\omega=\omega^\prime$ \footnote{The general TR-KMS generator in Eq.~\eqref{generator} allows a separate time-reversal-even Hamiltonian term (N) commuting with $H$. The continuous value of $h^K$ already reproduces the resonant Redfield coherent correction. Exact matching of the full generator therefore requires $N$}. 

We consider a factorized form for the dissipation kernel $K^K$. Then, the coefficients $K^K$ and $h^K$
are determined by the TR-KMS relation together with the above resonance matching.
This is specified next.

\subsection{Single channel} 
We begin with the single-channel case, where the interaction between the system and the bath is
$V=A\otimes B$, in product form of the system operator $A$ and the bath operator $B$. The coefficient kernels are scalar-valued, so
the channel indices in Eq.~\eqref{eq:completion-ansatz} are omitted,
and $E=\pm1$ drops out of the detailed balance relations.

The Redfield dissipative kernel reads
\begin{equation}
    K^R(\omega^\prime,\omega) = \frac12[\gamma(\omega)+ \gamma(\omega^\prime)] +\ui[ S(\omega)- S(\omega^\prime)],
\end{equation}
where $\gamma(\omega)$ and $S(\omega)$ are real functions, and $\gamma$ satisfies
\begin{equation}\label{eq:gammaomega}
    \gamma(\omega) = \ue^{\beta\omega}\gamma(-\omega).
\end{equation}
The TR-KMS
relations~\eqref{TR-KMSe1} and \eqref{TR-KMSe2} read
\begin{equation}\label{dbs1}
\begin{aligned}
     K(\omega^\prime,\omega) =& \ue^{\beta(\omega+\omega^\prime)/2}K(-\omega^\prime,-\omega) ,\\
    h(\omega',\omega)    =&
       \frac{\mathrm{i}\ue^{\beta(\omega+\omega')/2}}
     {2\sinh(\beta\nu/2)}
K(-\omega,-\omega')
\\&-
\frac{\mathrm{i}}{2}
\coth\!\left(\frac{\beta\nu}{2}\right)
K(\omega',\omega).
\end{aligned}
\end{equation}
We have written down $f[K]$ explicitly in the second equation.

We take a factorized form for the dissipation kernel $K^K$ and fix
its accompanying coherent coefficient $h^K$ through the second relation in TR-KMS
\begin{equation}\label{eq:single-completion-coefficients}
 \begin{aligned}
  K^K(\omega',\omega)&=q(\omega')^*q(\omega),\\
  h^K(\omega',\omega)
   &=f[K^K](\omega^\prime,\omega).
 \end{aligned}
\end{equation}
This factorized $K^K$ is physically favorable, 
as in the current single-channel case, it
allows the dissipative part to be expressed using a single Lindblad
jump operator $L$,
\begin{equation}
    \sum_{\omega^\prime\omega} q(\omega^\prime)^* q(\omega) \mathcal D_{\omega^\prime\omega}[\rho] =
    L \rho L^\dagger -\frac{1}2 \pb{L^\dagger L}{\rho},
\end{equation}
with $L= \sum_{\omega\in\Omega} q(\omega)A(\omega)$. 
Moreover, the master equation is automatically completely positive. 
We further decompose $q(\omega)$ in polar form, 
\begin{equation}
    q(\omega) = \ue^{\ui \phi(\omega)} r(\omega) .
\end{equation}
The match of $K^K(\omega,\omega)$ in Eq.~\eqref{resonance_matching} fixes $r(\omega)=\sqrt{\gamma(\omega)}$. 

Now we have coefficient 
\begin{equation}
    K^K(\omega^\prime,\omega) = \sqrt{\gamma(\omega)\gamma(\omega^\prime)}\ue^{\ui\phi(\omega)-\ui\phi(\omega^\prime)},
\end{equation}
where the modulus part $\sqrt{\gamma(\omega)\gamma(\omega^\prime)}$ 
is a geometric average, replacing the original arithmetic average. 
The detailed balance condition in Eq.~\eqref{dbs1} is fulfilled as long as $\phi(\omega)$ 
is an even function.

We have accomplished the detailed balance and the matching of $K^K$, 
leaving only the resonance matching condition of $h^K(\omega,\omega)=h^R(\omega,\omega)$, 
which will determine the phase $\phi(\omega)$. 
We now evaluate the continuous limit of $\omega^\prime\to\omega$ of $h^K$ in Eq.~\eqref{eq:single-completion-coefficients}. 
We fix $\omega$ and set $\omega'=\omega-\nu$. Expanding the modulus and
phase of the factorized kernel gives
\begin{equation*}
 \begin{aligned}
 K^K(\omega-\nu,\omega)
 =\gamma(\omega)-\frac{\nu}{2}\gamma'(\omega)+\ui\nu\gamma(\omega)\phi'(\omega)+O(\nu^2).
 \end{aligned}
\end{equation*}
By using the explicit expression of $h^K=f[K^K]$, the equal-frequency limit is obtained as
\begin{equation}\label{eq:single-hF-resonance}
 \begin{aligned}
  h^K(\omega,\omega)
    &=\lim_{\nu\to0}h^K(\omega-\nu,\omega)=\frac{\gamma(\omega)}{\beta}
      \bigl[\phi'(\omega)-\phi'(-\omega)\bigr].
 \end{aligned}
\end{equation}
We recall $\phi(\omega)$ should be an even function, so the expression in the square bracket is equal to $2\phi'(\omega)$.
Matching this to the Redfield value $h^R(\omega,\omega)$ in
Eq.~\eqref{eq:reference-hR-resonance} therefore fixes the derivative of phase:
\begin{equation}
    \phi^\prime (\omega) = \frac{S^\prime(\omega)-\ue^{\beta\omega}S^\prime(-\omega)}{2\gamma(\omega)}.
\end{equation}
    The remaining freedom of the integral constant does not affect the final dynamics, 
and we may choose
\begin{equation}
    \phi(\omega)=\int_0^\omega\ud \mu \frac{S^\prime(\mu)-\ue^{\beta\mu}S^\prime(-\mu)}{2\gamma(\mu)}.
\end{equation}
Noticing $\gamma(\omega)=\gamma(-\omega)\ue^{\beta\omega}$, 
it is easy to verify that the above solution for $\phi(\omega)$ is indeed an even function.

Under the resonance matching condition and the factorized form of the dissipation kernel, we have fixed all coefficients in the generator for the single channel.
The resulting TR-KMS completion master equation is
\begin{equation}\label{eq:single-completion-master}
  \dot\rho=-\ui[H+\lambda^2G^K,\rho]+\lambda^2\left(L\rho L^\dagger
                  -\frac12\pb{L^\dagger L}{\rho}\right),
\end{equation}
where
\begin{equation}\label{eq:single-completion-operators}
 \begin{aligned}
  L&=\sum_{\omega\in\Omega}\sqrt{\gamma(\omega)}
                       \ue^{\ui\phi(\omega)}A(\omega),\\
  G^K&=\sum_{\omega'\omega\in\Omega}h^K(\omega',\omega)
                               A(\omega')^\dagger A(\omega).
 \end{aligned}
\end{equation}
Here, $h^K$ is given by
Eq.~\eqref{eq:single-completion-coefficients}, including the continuous limit
\eqref{eq:single-hF-resonance}. The equation is completely positive and
satisfies TR-KMS detailed balance with respect to $\sigma$.
The modulus $\sqrt{\gamma(\omega)}$ in $L$ has the same spectral form as in geometric-arithmetic master equations, including the universal Lindblad equation~\cite{NathanRudner2020,NathanRudner2024,Davidovic2020,Davidovic2022}.
Nevertheless, those constructions do not contain our phase term $\phi(\omega)$, 
which is fixed by TR-KMS detailed balance and coherent resonance matching.

We remark that using the mean-force corrected $H$ instead of bare Hamiltonian $H_0$ indeed makes a difference in our matching strategy.
If the bare Hamiltonian $H_0$ is chosen as the reference Hamiltonian, the coherent coefficient that is waiting to be matched is the Lamb-shift coefficient $r^\text{LS}$.
According to Eqs.~\eqref{eq:gammaomega}, \eqref{eq:single-hF-resonance} and the fact that $\phi(\omega)$  is an even function, the continuous resonance limit of the TR-KMS coherent coefficient always satisfies
\begin{equation}\label{eq:single-reference-compatibility}
 h^K(\omega,\omega)+h^K(-\omega,-\omega)\ue^{\beta \omega}=0.
\end{equation}
However, without the $-g(\omega',\omega)$ term from mean-force correction in Eq.~\eqref{eq:reference-hR},
$r^\text{LS}(\omega,\omega)$ does not fulfill the above structure in general, 
since $r^\text{LS}(\omega,\omega)=S(\omega)$ and 
$   S(\omega)+e^{\beta \omega} S(-\omega)=-J(\omega)$ (from Eq.~\eqref{eq:hmf-J}), which is generally nonzero.
Thus, $h^K(\omega,\omega)$ cannot be matched to $r^\text{LS}(\omega,\omega)$.
Continuous resonance matching is generally incompatible with the bare reference in the present construction.

\subsection{Multiple channels}

We now apply Eq.~\eqref{eq:completion-ansatz} to
multiple coupling channels, $V=\sum_{i=1}^n A_i\otimes B_i$.
The coefficients $K^K$ and $h^K$ are now matrices on the channel
indices.
We require the coefficients to satisfy both
resonance matching, Eq.~\eqref{resonance_matching}, and TR-KMS
detailed balance, Eqs.~\eqref{TR-KMSe1} and \eqref{TR-KMSe2}.

We retain the factorized construction for the dissipation kernel $K^K$ and introduce the coherent coefficients from the second relation in TR-KMS detailed balance,
\begin{equation}\label{eq:multi-completion-coefficients}
 \begin{aligned}
  &K^K(\omega',\omega)=Q(\omega')^\dagger Q(\omega),\\
  &h^K(\omega',\omega)
   =f[K^K](\omega',\omega)
.
 \end{aligned}
\end{equation}
Here $Q(\omega)$ is an $n\times n$ matrix for
$n$ coupling channels. The value of $h^K(\omega,\omega)$
at equal frequencies is defined by continuity.
The factorization guarantees complete positivity
and allows at most $n$ Lindblad jump operators,
\begin{equation}
    L_k=\sum_{j;\omega\in\Omega} Q_{kj}(\omega) A_j(\omega)
\end{equation}
The indices $i,j$ denote coupling channels,
while $k$ labels the jump operators.

Similar to the single-channel case, we separate the
spectral magnitude from the phase through the polar decomposition in matrix form
\begin{equation}\label{eq:multi-polar}
 Q(\omega)=U(\omega)R(\omega),
\end{equation}
where $R$ is a positive Hermitian matrix and $U$ is a unitary matrix.
Now the dissipation kernel reads 
\begin{equation}\label{eq:multi-polar-kernel}
 K^K(\omega',\omega)
   =R(\omega')U(\omega')^\dagger U(\omega)R(\omega)
\end{equation}
At resonance, $\omega=\omega^\prime$, $U$ drops out of the kernel $K^K$, so dissipative matching
fixes $R$, as a matrix square root,
\begin{equation}\label{eq:multi-magnitude-matching}
 \begin{aligned}
  R(\omega)&=\sqrt{\gamma(\omega)}.
 \end{aligned}
\end{equation}
With the bath symmetries in
Eq.~\eqref{gamma_r}, $\gamma(\omega) =\ue^{\beta\omega} E\gamma(-\omega)E^T$, 
we find that the dissipative TR-KMS relation \eqref{TR-KMSe1} is fulfilled 
as long as the phase matrix obeys
\begin{equation}\label{eq:multi-U-symmetry}
 U(\omega)=E U(-\omega)E^T,
\end{equation}
which serves as an analogue of the even function requirement in the single-channel case. 

The remaining task is to match the coherent coefficients $h^K(\omega,\omega)$ and $h^R(\omega,\omega)$ at 
resonance.
To do so, we now evaluate $h^K(\omega,\omega)$ as the continuous
limit of $\omega^\prime\to\omega$ in Eq.~\eqref{eq:multi-completion-coefficients}. 
The detailed procedure of this limit is shown in Appendix~\ref{app:completion-details}, 
and the final result reads
\begin{equation}\label{eq:continuous-hK}
    h^K(\omega,\omega)
=-\frac{i}{\beta}\bigl[\Xi(\omega)+E\Xi(\omega)^TE^T\bigr],
\end{equation}
with an anti-Hermitian matrix
\begin{equation}
      \Xi=R U^\dagger U'R+\frac12[R,R'].
\end{equation}
All matrices are evaluated at
$\omega$, and primes denote frequency derivatives.
The commutator term $[R,R']$ accounts for the
noncommutativity of the frequency-dependent matrix magnitude, 
which vanishes in the single-channel case.

On the other hand, the Redfield target from Eq.~\eqref{eq:reference-hR-resonance} is  
\begin{equation}
    h^R(\omega,\omega)=\frac{D(\omega)}{\beta},\quad D(\omega)=S'(\omega)-e^{\beta\omega}S'(-\omega)^T.
\end{equation}
The matching condition $h^K(\omega,\omega)=h^R(\omega,\omega)$ then gives 
\begin{equation}
    \Xi+E\Xi^TE^T=iD .
\end{equation}
The solution to the above equation is not unique. 
From Eq.~\eqref{gamma_r}, one can verify that $E D^T E^T=D$, and therefore $\Xi=\ui D/2$ is one solution. 
In general, we can write
$\Xi=\ui D/2+Y$, 
with an additional term $Y$ satisfying the following equations,
\begin{equation}\label{eq:multi-Y-constraints}
 \begin{aligned}
  Y&=-Y^\dagger,\quad
  EYE^T=-Y^T.
 \end{aligned}
\end{equation}
Then, writing down the expression of $\Xi$ yields the matrix phase differential equation
\begin{equation}\label{eq:multi-phase-equation}
 U^\dagger U'
   =R^{-1}\left(\frac\ui2D+Y-\frac12[R,R']\right)R^{-1}.
\end{equation}
We are able to solve the matrix $U$ from the above differential equation, 
with a specific initial condition $U(0)=I$. 
Other initial condition for $U(0)$ does not change the corresponding dynamics.
The right-hand side is anti-Hermitian, preserving the unitarity of matrix $U$.

The last step is to ensure the matrix $U$ obtained from the above differential equation indeed satisfies the frequency-reversal symmetry in Eq.~\eqref{eq:multi-U-symmetry}, which comes from the first relation of the TR-KMS condition.
This requirement gives another constraint on $Y$, 
\begin{equation}
  Y(\omega)=-\ue^{\beta\omega}E Y(-\omega)E^T.
\end{equation}
The derivation is given in Appendix \ref{app:completion-symmetry}.
We note that $Y=0$ fulfills all the constraints on $Y$. 

Different allowed choices of $Y(\omega)$
change the cross-frequency kernel. We select
the canonical completion by setting $Y=0$. 
The physical motivation is to preserve the additivity in the case of mutually independent bath channels. 
This $Y=0$ choice also minimizes the first-order
near-resonant normalized deviation of the dissipative kernel from Redfield.
Please see Appendix~\ref{app:canonical choice} for a detailed discussion.

With this choice, the differential equation
then uniquely determines $U(\omega)$ from the differential equation
\begin{equation}\label{canonical}
    U^\dagger U'    =R^{-1}\left(\frac\ui2D-\frac12[R,R']\right)R^{-1},
\end{equation}
and an initial condition $U(0)=I$. 
The resulting multichannel TR-KMS completion master
equation is
\begin{equation}\label{eq:multi-completion-master}
 \begin{aligned}
  \dot\rho={}&-\ui[H+\lambda^2G^K,\rho]\\
   &+\lambda^2\sum_{k=1}^n\left(L_k\rho L_k^\dagger
                       -\frac12\pb{L_k^\dagger L_k}{\rho}\right),
 \end{aligned}
\end{equation}
where
\begin{equation}\label{eq:multi-completion-operators}
 \begin{aligned}
  L_k&=\sum_{j;\omega\in\Omega}[U(\omega)R(\omega)]_{kj}A_j(\omega),\\
  G^K&=\sum_{ij;\omega'\omega\in\Omega}h^K_{ij}(\omega',\omega)
                              A_i(\omega')^\dagger A_j(\omega).
 \end{aligned}
\end{equation}
Here, $R=\sqrt\gamma$, and $U$
is determined by the differential equation above. The coherent
coefficients are given by Eq.~\eqref{eq:multi-completion-coefficients}. The equation is completely
positive, satisfies TR-KMS detailed balance with respect to
$\sigma$, and retains interference between distinct Bohr frequencies.

\subsection{Problem with nTR-KMS}
Our resonance matching procedures reveal the distinction between
TR-KMS detailed balance and nTR-KMS detailed balance.
If we impose nTR-KMS detailed balance, the completion scheme generally has
no solution.
Here, we show how this distinction appears in the case of single channel.

We recall that each detailed balance imposes two relations on the coefficients.
The TR-KMS and nTR-KMS differ only in the first relation, 
which constrains $K(\omega^\prime,\omega)$ with $K(-\omega^\prime,-\omega)$; see Eqs.~\eqref{TR-KMSe1} and \eqref{eq:db-ntr-coefficients1}.
For the second relation, both TR-KMS and nTR-KMS impose the same requirement for $h(\omega^\prime,\omega)$ if $K(\omega^\prime,\omega)$ is given; see Eqs.~\eqref{TR-KMSe2} and \eqref{eq:db-ntr-coefficients2}.

If we use nTR-KMS detailed balance in our construction, the procedures up to Eq.~\eqref{eq:single-hF-resonance} can be made similar. 
We still take a factorized dissipation kernel
and write it down in polar form, 
\begin{equation}
    K^N(\omega^\prime,\omega) = \ue^{\ui[\phi(\omega)-\phi(\omega^\prime)]}r(\omega^\prime)r(\omega).
\end{equation}
The coherent correction is still defined through the detailed balance condition, 
\begin{equation}\label{nTR-construction-h}
    h^N(\omega^\prime,\omega) = f[K^N](\omega^\prime,\omega),
\end{equation}
which is shared by both TR-KMS and nTR-KMS. 
Here, we have used $K^N$ and $h^N$ in nTR-KMS to distinguish it from the TR-KMS.
The matching between $K^N(\omega^\prime,\omega)$ and Redfield coefficient $K^R(\omega^\prime,\omega)$ still gives $r(\omega)=\sqrt{\gamma(\omega)}$ for nTR-KMS. 
The difference lies in the property on the phase $\phi(\omega)$:
imposing the first relation of nTR-KMS \eqref{eq:db-ntr-coefficients1} yields that $\phi(\omega)$ should be an odd function (up to an overall constant $c$), $\phi(\omega)=-\phi(-\omega)+c$. 
This is in contrast to TR-KMS, where $\phi(\omega)$ should be an even function in $\omega$. 

According to Eq.~\eqref{nTR-construction-h}, 
when taking the continuous limit of $\omega^\prime\to\omega$, 
we still have the following expression for $h^N$ at the resonance, same as Eq.~\eqref{eq:single-hF-resonance},
$$ h^N(\omega,\omega)
 =\frac{\gamma(\omega)}{\beta}
      [\phi'(\omega)-\phi'(-\omega)].
$$
For nTR-KMS detailed balance, the phase $\phi(\omega)$ should be an odd function instead of even function, so that the above $h^N(\omega,\omega)=0$ must vanish. 
Therefore, nTR-KMS does not allow any nonzero continuous coherent correction at all.
On the other hand, the Redfield coefficient at resonance $h^R(\omega,\omega)$ in Eq.~\eqref{eq:reference-hR-resonance} is usually nonzero. 
The continuous matching requires $h^N(\omega,\omega)$ and $h^R(\omega,\omega)$ to be equal, 
and this can never be achieved. 

The incompatibility of the nTR-KMS shows that, 
under our construction scheme, the nTR-KMS detailed balance cannot be restored while keeping the coefficients close to the Redfield equation.
In the next section on error analysis, we show that the resonance matching scheme ensures 
that the constructed TR-KMS completion dynamics has a second-order accuracy bound, 
while this precision is not yet achieved for any nTR-KMS dynamics. 
It indicates the TR-KMS detailed balance could be more appropriate to construct dynamics of open quantum systems that faithfully reproduce microscopic evolution.

\section{Accuracy relative to microscopic dynamics}\label{sec:error}

In this section, we consider a fixed finite-dimensional system and analyze the error of the
above TR-KMS completion dynamics relative to the microscopic
reduced dynamics. Based on the two conditions on the 
error of the Redfield approximation and rapid mixing, we
obtain a time-uniform $O(\lambda^2)$ error bound between the
TR-KMS completion state $\rho_K(t)=\ue^{t\mathcal L^K}\rho_0$ and the
microscopic reduced state $\rho_{\rm mic}(t)$. 

The derivation has two steps. First, we introduce an
auxiliary comparison state that differs from the physical state by
only $O(\lambda^2)$. Resonance matching cancels the second-order
generator difference in its evolution equation, leaving a
continuously generated error of order $\lambda^4$.
Second, if perturbations decay on the relaxation timescale
$O(\lambda^{-2})$, past errors are damped rather than accumulating
indefinitely, giving a total state error of $O(\lambda^2)$.

We use the trace norm $\norm{\cdot}_1$ for both state errors and
error sources throughout this section.

Our analysis is based on two physical assumptions. 
\emph{(i) Controlled microscopic Redfield error.} The microscopic exact evolution
satisfies
\begin{equation}\label{eq:error-micro}
 \dot\rho_{\rm mic}=\mathcal L^R\rho_{\rm mic}
       +\lambda^4r_\lambda(t)+\lambda^2 b_\lambda(t),
\end{equation}
where $\sup_t\norm{r_\lambda(t)}_1\leq C_r$ and
$\int_0^\infty\norm{b_\lambda(t)}_1\ud t\leq C_b$.
Here $C_r$ and $C_b$ are independent of time and
sufficiently small $\lambda$.
The persistent source $\lambda^4r_\lambda$ is
fourth order, while the initial-memory transient
$\lambda^2b_\lambda$ is second order but decays in time.

\emph{(ii) Rapid mixing condition.}
Differences between two initial density matrices $\rho_1$ and
$\rho_2$ decay under the TR-KMS completion dynamics according to
\begin{equation}\label{eq:error-mixing}
 \begin{gathered}
 \norm{\ue^{u\mathcal L^K}(\rho_1-\rho_2)}_1
 \leq a_\lambda(u)\norm{\rho_1-\rho_2}_1,\\
 \int_0^\infty a_\lambda(u)\ud u
 \leq \frac{A_{\rm mix}}{\lambda^2}.
 \end{gathered}
\end{equation}
Here $a_\lambda(u)\geq0$, and $A_{\rm mix}$ is
independent of time and small $\lambda$. The integral bound states
that the effective lifetime of a perturbation is at most
$O(\lambda^{-2})$. 
Exponential relaxation
$a_\lambda(u)=M_{\rm mix}\ue^{-g\lambda^2u}$ with fixed
$M_{\rm mix},g>0$ is sufficient for this condition.
We note that in the above formula, $\rho_1-\rho_2$ can be replaced by any traceless operator.

For the finite-dimensional construction with the regular coefficient
functions specified above, assumptions (i) and (ii) yield the
time-uniform $O(\lambda^2)$ error bound derived below.

We expand the Redfield generator $\mathcal L^R$
and the TR-KMS completion generator $\mathcal L^K$
using the bare Bohr operators $A_i^0(\omega)$ of $H_0$,
\begin{equation}\label{eq:error-reference-remainder}
 \begin{aligned}
  \mathcal L^R&=\mathcal L_0+\lambda^2\mathcal L_2^R,\\
  \mathcal L^K&=\mathcal L_0+\lambda^2\mathcal L_2^K
                            +\lambda^4\mathcal E_\lambda,
 \end{aligned}
\end{equation}
with the uncoupled evolution $\mathcal L_0=-\ui\operatorname{ad}_{H_0}$. 
The second-order terms are explicitly
\begin{equation}\label{eq:error-second-order-generators}
 \begin{aligned}
  \mathcal L_2^R={}&-\ui\operatorname{ad}_{H_2^{\rm MF}}\\
  &+\sum_{ij;\omega'\omega\in\Omega_0}\left[
    K^R_{ij}\mathcal D^0_{ij;\omega'\omega}
    -\ui h^R_{ij}\operatorname{ad}_{P^0_{ij;\omega'\omega}}
    \right],\\
  \mathcal L_2^K={}&-\ui\operatorname{ad}_{H_2^{\rm MF}}\\
  &+\sum_{ij;\omega'\omega\in\Omega_0}\left[
    K^K_{ij}\mathcal D^0_{ij;\omega'\omega}
    -\ui h^K_{ij}\operatorname{ad}_{P^0_{ij;\omega'\omega}}
    \right].
 \end{aligned}
\end{equation}
All coefficient functions are evaluated at the
bare Bohr frequencies $(\omega',\omega)$. The
first expression reproduces the standard Redfield equation
\eqref{eq:reference-Redfield}. The second uses bare operators $A^0_i(\omega)$
in place of $A_i(\omega)$ in the TR-KMS completion
equation \eqref{eq:multi-completion-master}, while retaining the
coefficient functions $K^K$, $h^K$, and $H^{\rm ref}$.
This induces a $O(\lambda^4)$ error in the generator, collected in $\lambda^4\mathcal E_\lambda$.

We define the coefficient differences between two generators, 
$\delta K=K^K-K^R$ and $\delta h=h^K-h^R$.
Resonance matching in Eq.~\eqref{resonance_matching} makes them vanish with $O(\nu)$ in the limit $\nu=\omega-\omega'\to0$, 
allowing us to define a superoperator
\begin{equation}\label{eq:error-M}
 \mathcal W=\ui\sum_{ij\omega'\omega}\left[
 \frac{\delta K_{ij}}{\nu}\mathcal D^0_{ij;\omega'\omega}
 -\ui\frac{\delta h_{ij}}{\nu}\operatorname{ad}_{P^0_{ij;\omega'\omega}}
 \right].
\end{equation}
Due to the resonance matching, 
its value at $\nu=0$ is well defined through continuity.
With $\mathcal L_0 A^0_i(\omega) = \ui \omega A^0_i(\omega)$, 
we can verify the following relation, which rewrites the difference 
between two second-order generators by using $\mathcal W$,
\begin{equation}\label{eq:error-commutator}
 \mathcal L_2^K-\mathcal L_2^R=[\mathcal W,\mathcal L_0].
\end{equation}
Here the brackets denote commutators under
superoperator composition.

To compare the two dynamics, we introduce a dressing
transformation $\mathcal B_\lambda=I+\lambda^2\mathcal W$.
Similar dressing transformations appear in the derivation and
error analysis of the ULE~\cite{NathanRudner2020,NathanRudner2024}.
This transformation
cancels the second-order difference between the Redfield and
TR-KMS completion generators:
\begin{equation}\label{eq:error-intertwining}
 \mathcal B_\lambda\mathcal L^R-\mathcal L^K\mathcal B_\lambda
 =\lambda^4\mathcal Q_\lambda.
\end{equation}
The remainder is explicitly given by
\begin{equation}\label{eq:error-Q}
 \mathcal Q_\lambda
  =\mathcal W\mathcal L_2^R-\mathcal L_2^K\mathcal W
                         -\mathcal E_\lambda\mathcal B_\lambda.
\end{equation}
Applying $\mathcal B_\lambda$ to $\rho_{\rm mic}$ adds only a bounded $O(\lambda^2)$
correction to the state. 
In the following, we will compare the TR-KMS completion state $\rho_K$ with 
the dressed microscopic state $\mathcal B_\lambda\rho_{\rm mic}$.

We set $z=\mathcal B_\lambda\rho_{\rm mic}-\rho_K$. 
Taking its time derivative and 
using Eqs.~\eqref{eq:error-micro} and \eqref{eq:error-intertwining} imply
\begin{equation}\label{eq:error-z-evolution}
 \dot z=\mathcal L^K z+\lambda^4\mathcal F_\lambda(t)
                    +\lambda^2\widetilde b_\lambda(t),
\end{equation}
where $\mathcal F_\lambda=\mathcal Q_\lambda\rho_{\rm mic}
+\mathcal B_\lambda r_\lambda$ and
$\widetilde b_\lambda=\mathcal B_\lambda b_\lambda$.
These sources have zero trace and do not change the
normalization of the state.

We investigate the error bound on $\mathcal F_\lambda=\mathcal Q_\lambda\rho_{\rm mic}
+\mathcal B_\lambda r_\lambda$ and $\widetilde b_\lambda=\mathcal B_\lambda b_\lambda$.
For a finite-dimensional system, superoperators 
$\mathcal W$, $\mathcal E_\lambda$, $\mathcal L^R_2$, $\mathcal L^K_2$ are all 
bounded, which ensures that $\mathcal W\rho_{\rm mic}$ and $\mathcal Q_\lambda\rho_{\rm mic}$ 
remain bounded at all time. 
Together with the assumption (i) on $r_\lambda$ and $b_\lambda$,
we know that $\mathcal F_\lambda$ is uniformly bounded in time, and
$\widetilde b_\lambda$ is integrable in time.
We denote these bounds by
\begin{equation*}
 \begin{aligned}
  \sup_{t\geq0}\norm{\mathcal W\rho_{\rm mic}(t)}_1
   &\leq C_W,\quad
  \sup_{t\geq0}\norm{\mathcal F_\lambda(t)}_1
   \leq C_F,\\
  \int_0^\infty\norm{\widetilde b_\lambda(t)}_1\ud t
   &\leq\widetilde C_b.
 \end{aligned}
\end{equation*}
Here three constants, $C_W$, $C_F$, and $\widetilde{C}_b$, are independent of time and
small $\lambda$.

With $z(0)=\lambda^2\mathcal W\rho_0$, integrating the differential equation \eqref{eq:error-z-evolution} yields
\begin{equation}\label{eq:error-duhamel}
 \begin{aligned}
 z(t)={}&\lambda^2 \ue^{t\mathcal L^K}\mathcal W\rho_0
 +\lambda^2\int_0^t\ud s\,\ue^{(t-s)\mathcal L^K}\widetilde b_\lambda(s)\\
 &
 +\lambda^4\int_0^t\ud s\,\ue^{(t-s)\mathcal L^K}\mathcal F_\lambda(s).
 \end{aligned}
\end{equation}
On the right-hand side, the first term and
the second term contribute at most $\lambda^2C_W$ and $\lambda^2\widetilde C_b$, respectively. 
For the third term, the mixing condition (ii) can be applied to
the traceless operator $\mathcal F_\lambda(s)$,
which bounds its remaining contribution by
$\lambda^4C_F\int_0^t a_\lambda(u)\ud u\leq \lambda^2 C_F A_{\rm mix}$ \cite{ikeuchi2026}.
Finally, returning to the physical states gives
\begin{equation}
    \rho_{\rm mic}(t)-\rho_K(t)=z(t)-\lambda^2\mathcal W\rho_{\rm mic}(t),
\end{equation}
where the last term contributes at most $\lambda^2C_W$ to the error bound. 
Adding these contributions yields
the time-uniform bound
\begin{equation}\label{eq:error-uniform}
\sup_{t\geq0}\norm{\rho_{\rm mic}(t)-\rho_K(t)}_1
 \leq\lambda^2(2C_W+\widetilde C_b+C_FA_{\rm mix}).
\end{equation}
This is our main result on long-time accuracy:
the TR-KMS completion dynamics approximates
the microscopic reduced dynamics with a time-uniform
$O(\lambda^2)$ error.

We conclude with comments on the two assumptions.
For (i), time-uniform fourth-order Born--Markov remainders have been
proved for Gaussian thermal baths with finite first absolute time moments of the
bath correlations and bounded system coupling operators~\cite{ScandiAlhambra2026,AgerskovNathan2026}.
Gaussianity of the bath can be relieved to some extend in this derivation: multipoint clustering
also underlies weak-coupling derivations for more general baths~\cite{FacchiLigaboYuasa2017}.
A sufficient route to (i) for non-Gaussian even baths requires the relevant
connected multipoint responses to have integrable memory with
uniformly controlled higher-order contributions at weak coupling; short two-point memory alone is
insufficient.

Assumption (ii) concerns system relaxation in the TR-KMS completion dynamics: perturbations in density matrix must have an effective lifetime at most of order $\lambda^{-2}$.
This condition can be checked directly from the constructed generator
$\mathcal L^K$. It holds, for example, if $\mathcal L^K$ has a simple
zero eigenvalue and a dissipative gap of at least $\sim\lambda^2$, 
and if its eigenoperators form a basis
that remains well conditioned as $\lambda\to0$.
Without Assumption (ii), the trace-distance contraction in Eq.~\eqref{eq:error-duhamel} gives
\begin{equation}\label{eq:error-without-mixing}
 \norm{\rho_{\rm mic}(t)-\rho_K(t)}_1
 \leq\lambda^2(2C_W+\widetilde C_b)+\lambda^4C_Ft.
\end{equation}
Thus the state error remains $O(\lambda^2)$ throughout the dissipative
timescale $t=\tau_d\sim\lambda^{-2}$ even without mixing. 
Assumption (ii) extends this control to arbitrarily long times.

We remark that the present proof works for finite-dimensional system. 
Proving analogous bounds for broad classes
of physically relevant infinite-dimensional systems remains an open
problem.

\section{Benchmark with two harmonic oscillators and harmonic baths}
\label{sec:oscillator-benchmark}

In this section, we study two coupled near-resonant oscillators
interacting with harmonic baths across secular and nonsecular regimes.
We benchmark the TR-KMS completion, Redfield, Davies, SA, and
high-temperature Caldeira--Leggett master equations against the exact
microscopic dynamics.
The results confirm the predicted weak-coupling scalings: TR-KMS completion has an
$O(\lambda^2)$ error in the transient state and an $O(\lambda^4)$ steady-state error. Among
all the detailed-balance master equations compared here, it achieves the most favorable dynamical accuracy.

\begin{figure*}[t]
 \centering
 \includegraphics[width=\textwidth]{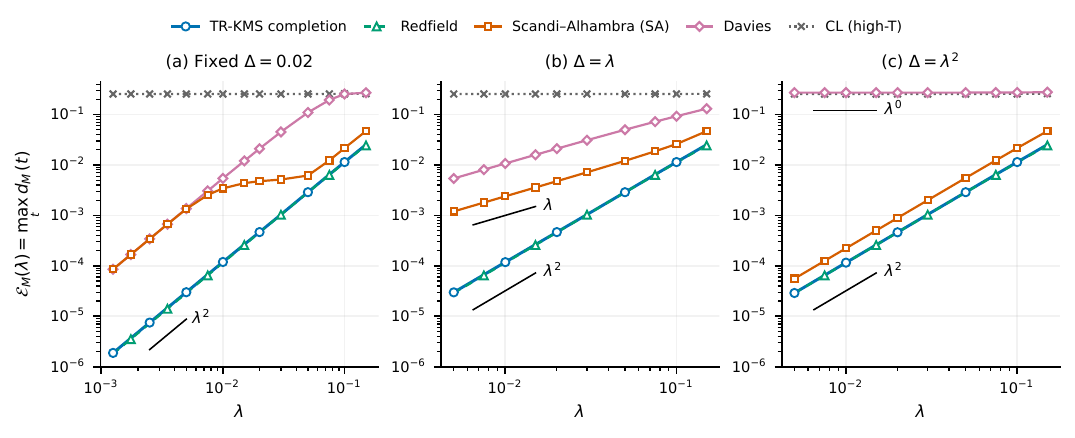}
 \caption{Dynamical error against the exact microscopic
 evolution for three regimes: (a) the secular regime with $\Delta=\text{const}$ (b) the intermediate regime with $\Delta=\lambda$ (c) the nonsecular regime with $\Delta=\lambda^2$. TR-KMS completion and Redfield
 retain second-order accuracy throughout. SA has a first-order error
 for $\Delta=\lambda$, while Davies has a finite error for
 $\Delta=\lambda^2$. For fixed nonzero splitting (a), all four recover
 second-order scaling at sufficiently weak coupling. The high-temperature CL approximation outside its temperature regime and has a finite error. Black segments indicate the asymptotic power laws.}
 \label{fig:bench-dynamics}
\end{figure*}

\subsection{Model and microscopic reference}
We consider two coupled harmonic oscillators, each interacting with an
independent thermal bath;  a
related oscillator model is studied in Ref.~\cite{RecabalEtAl2026}.
We set $\hbar=k_B=m=\omega_0=1$, where $\omega_0$
is the uncoupled oscillator frequency. The system Hamiltonian is
\begin{equation}\label{eq:bench-oscillator}
 H_{\rm osc}=\frac12\sum_{\alpha=1}^2(p_\alpha^2+q_\alpha^2)
                  +kq_1q_2.
\end{equation}
For small $|k|$, the two normal-mode frequencies
$\sqrt{1\pm k}$ are nearly resonant. We parameterize their separation by
\begin{equation}\label{eq:bench-gap}
 \Delta=\sqrt{1+k}-\sqrt{1-k}.
\end{equation}
The secular approximation requires
$\Delta\gg\lambda^2$. Otherwise, intermode
interference remains important over relaxation.

Each position couples to its own Caldeira--Leggett reservoir~\cite{CaldeiraLeggett1983,HsiangArisoyHu2022} 
with different coupling strengths $\lambda d_1$ and $\lambda d_2$. 
The total Hamiltonian reads 
\begin{equation}\label{eq:bench-microscopic}
 \begin{aligned}
 \tilde H={}&H_{\rm osc}+\sum_{\alpha\ell}\bigg[
 \frac{P_{\alpha\ell}^2}{2m_\ell}+\frac{m_\ell\omega_\ell^2}{2}\bigg(
 X_{\alpha\ell}
 -\frac{\lambda d_\alpha }
              {m_\ell\omega_\ell^2}q_\alpha\bigg)^2\bigg],
 \end{aligned}
\end{equation}
with $d_1=1$ and $d_2=1/\sqrt3$. 
Both baths have inverse temperature
$\beta=2$ and the same Ohmic spectrum with a double Lorentz--Drude cutoff,
\begin{equation}\label{eq:bench-spectrum}
 \begin{aligned}
 J_0(w)&=\frac{\pi}{2}\sum_\ell
             \frac{1}{m_\ell\omega_\ell}
                         \delta(w-\omega_\ell)\\
 &=\frac{w\Omega_c^2\Omega_h^2}
          {(w^2+\Omega_c^2)(w^2+\Omega_h^2)},\qquad w>0,
 \end{aligned}
\end{equation}
with two cutoff frequencies $\Omega_c=5$ and $\Omega_h=10$. The two cutoff
factors give a finite equal-time bath correlation while preserving an exponential decay of time-dependent bath correlation. The counterterm in Eq.~\eqref{eq:bench-microscopic} is retained in all comparisons.

Initially, the oscillators occupy the local coherent
state $|\alpha_1=1\rangle\otimes|0\rangle$, uncorrelated with the
thermal reservoirs. The quadratic microscopic Hamiltonian gives exact
Gaussian dynamics and an independently calculable mean-force Gibbs
state~\cite{HuPazZhang1992,FlemingRouraHu2011,FlemingRouraHuMultipartite2011}. These exact results provide a common
reference for evaluating the errors of TR-KMS completion, nonsecular
Redfield, fully secular Davies, the SA equation, and the high-temperature
Caldeira--Leggett (CL) equation.

In this model, all master equations can be treated analytically.
For the SA equation, we choose the temporal coarse-graining
window as $T_{\rm cg}=1/\lambda$, following the proposal in Ref.~\cite{ScandiAlhambra2026}.
The high-temperature approximation underlying the CL equation is not satisfied here, so it is included only as an applicability check.

We briefly illustrate how our TR-KMS completion dynamics is obtained here. 
We first obtain $H^{\rm ref}$ from the second order expansion of Hamiltonian of mean force and then compute its Bohr components at the positive and negative mode frequencies.
Since the baths are independent, we can apply the single-channel
formulas~\eqref{eq:single-completion-coefficients}\eqref{eq:single-completion-operators} to each bath to obtain coefficients and directly add the resulting coherent corrections and dissipators. 
Both channels use the same frequency-domain bath correlation functions, $\gamma_{\rm b}/2$ and $S_{\rm b}$ (Eq. \eqref{eq:prelim-bath-transform}), which can be computed from the spectrum $J_0$ of the bath.

\subsection{Accuracy comparison}
Here we test the accuracy for the above-mentioned master equations as
the coupling strength $\lambda$ goes to zero.
We are interested in the nonsecular regime,
where the frequency difference $\Delta$ is comparable
to the dissipative strength $\lambda^2$.
Therefore, we consider three different scalings for $\Delta$ in the limit of $\lambda\to0$:
(a) $\Delta=\text{const}$, (b) $\Delta\sim\lambda$, (c) $\Delta\sim \lambda^2$.

In the first case, we set fixed $\Delta=0.02$. The secular approximation is valid at sufficiently weak coupling.
In the second case, we simply take $\Delta=\lambda$.
The interference between the two modes persists at the timescale of the coarse-graining time window $T_{\rm cg}\sim1/\lambda$ for the SA master equation in Ref.~\cite{ScandiAlhambra2026}
but its accumulated contribution on the dissipation
timescale $\tau_d\sim\lambda^{-2}$ is suppressed by rapid oscillations.
We call this case the intermediate regime.
In the third case, we take $\Delta=\lambda^2$. The quantum interference persists over dissipation time ($\tau_d\sim\lambda^{-2}$), and the secular approximation is not justified on this timescale.
In the latter two cases, we change the oscillator coupling $k$ with $\lambda$ to maintain the specified frequency relation.

For each equation, we measure the error against the
exact microscopic reduced state and define
the maximum error during the whole evolution $\mathcal E_M(\lambda)$,
\begin{equation}\label{eq:bench-error}
 \begin{aligned}
 d_M(t)&=\frac12\norm{\rho_M(t)-\rho_{\rm mic}(t)}_1,\\
 \mathcal E_M(\lambda)&=\max_{t\geq0}d_M(t).
 \end{aligned}
\end{equation}
Our comparison focuses on this maximum error during the evolution, 
and regards it as the key quantity quantifying the 
predictability of different master equations.

This maximum error $\mathcal E_M(\lambda)$ of each equation is shown in
Fig.~\ref{fig:bench-dynamics}, with varying coupling strength $\lambda$. 
Three different regimes (a)-(c) correspond to Fig.~\ref{fig:bench-dynamics}(a)-(c) respectively. 
Since the high-temperature CL equation is out of its applicable range, 
it has a finite error which does not vanish in the weak-coupling limit. 
In the case of fixed $\Delta$, the other four master equations provide an error scaling as $O(\lambda^2)$ for sufficiently small $\lambda$.
This matches our expectation; since fixed $\Delta$ leads to the secular regime for small $\lambda$, these equations should be valid.
In the regime (b) with $\Delta=\lambda$, SA and Davies have first-order errors $O(\lambda)$, whereas our TR-KMS completion and Redfield retain second-order $O(\lambda^2)$. 
For regime (c) with $\Delta=\lambda^2$, SA recovers second-order scaling, but the Davies error approaches
a finite constant rather than vanishing. 
Our TR-KMS completion and Redfield retain second-order $O(\lambda^2)$. 
We summarize the concrete fitted powers in the three regimes in Table~\ref{tab:bench-powers}.

Our TR-KMS completion dynamics closely matches the dynamical accuracy of Redfield. 
Both equations retain an $O(\lambda^2)$ state-error scaling across all three regimes and yield the smallest errors among the equations compared here. 
This advantage in accuracy persists even when other master equations exhibit the same $O(\lambda^2)$ scaling of error, including the secular regime (a) and the SA equation in case (c). 
The estimation in this example justifies our completion: it indeed preserves second-order dynamical accuracy of Redfield while restoring complete positivity and TR-KMS detailed balance.

\begingroup
\makeatletter
\let\ftype@table\ftype@figure
\makeatother
\begin{table}[th]
 \caption{Weak-coupling powers fitted to
 $\mathcal E_M\propto\lambda^p$ in
 each regime. The final row gives the worst-case error order for general open systems.
 }
 \label{tab:bench-powers}
 \begin{ruledtabular}
 \begin{tabular}{lrrrr}
  & TR-KMS & Redfield & SA & Davies\\
 \hline
 $\Delta=0.02$       &1.999&2.000&1.999&1.999\\
 $\Delta=\lambda$    &1.987&2.000&0.994&0.984\\
 $\Delta=\lambda^2$  &1.999&2.000&1.999&0.000\\
 \hline
 Worst-case error
 &$O(\lambda^2)$
 &$O(\lambda^2)$
 &$O(\lambda)$
 &$O(1)$
 \end{tabular}
 \end{ruledtabular}
\end{table}
\endgroup

Table \ref{tab:bench-powers} summarizes the least favorable error scaling observed for each equation across the three frequency scans. 
Among three detailed-balance master equations, the worst-case dynamical errors scale as $O(1)$ for Davies, $O(\lambda)$ for SA, and $O(\lambda^2)$ for TR-KMS completion.
The least favorable error happens in different regimes. 
Davies exhibits its worst scaling at nonsecular regime ($\Delta=\lambda^2$) while SA exhibits its worst scaling at intermediate regime ($\Delta=\lambda$). On the other hand, TR-KMS completion exhibits second-order scaling in all three regimes.

These three regimes can coexist in a dense many-body spectrum, where different pairs of transitions can have appreciable coupling weights at different frequency separations. 
In a real many-body system with quasi-continuous spectrum, 
there are a large number of close frequency differences, 
and the set of frequency differences $\{\nu\}_{ij} =\omega_i-\omega_j$ may span across regions from $\lambda^2$ to $\lambda^0$, covering all three regimes (a)-(c). 
The least favorable error scalings observed in the oscillator benchmark can therefore be attained simultaneously in a complex system \footnote{A single composite system can be constructed in which all three equations exhibit their respective worst-case error scalings.}.
On the dissipative timescale, the oscillator benchmark reproduces
the weak-coupling error scalings summarized in Table~\ref{tab:comparison}.

\begin{figure}[t]
 \centering
 \includegraphics[width=\columnwidth]{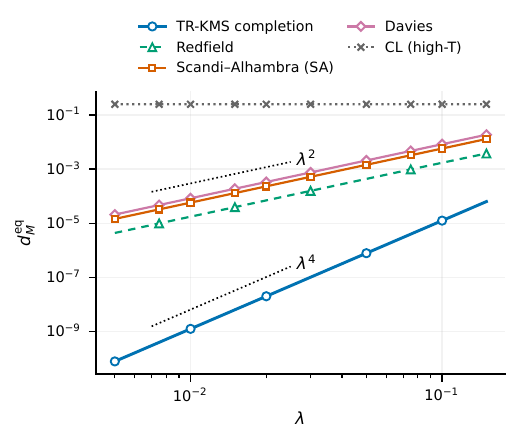}
 \caption{Steady-state error relative to the
 exact mean-force state for
 $\Delta=\lambda$. TR-KMS completion reproduces its second-order
 correction, leaving an $O(\lambda^4)$ error. Redfield, SA, and Davies
 have $O(\lambda^2)$ errors. Davies can therefore approach the correct
 weak-coupling equilibrium while inaccurately describing relaxation.
 The high-temperature CL approximation retains a finite error at this
 temperature.}
 \label{fig:bench-equilibrium}
\end{figure}

We also compare the steady-state errors for $\Delta=\lambda$
in Fig.~\ref{fig:bench-equilibrium}.
It shows a fourth-order steady-state error
for TR-KMS completion, compared with second-order errors for Redfield,
SA, and Davies. This confirms that the TR-KMS completion
provides a fourth order error for the equilibrium state, $d_K(t=+\infty)\sim\lambda^4$.

\subsection{Evoluation of the energy error}
\begin{figure*}[t]
 \centering
 \includegraphics[width=\textwidth]{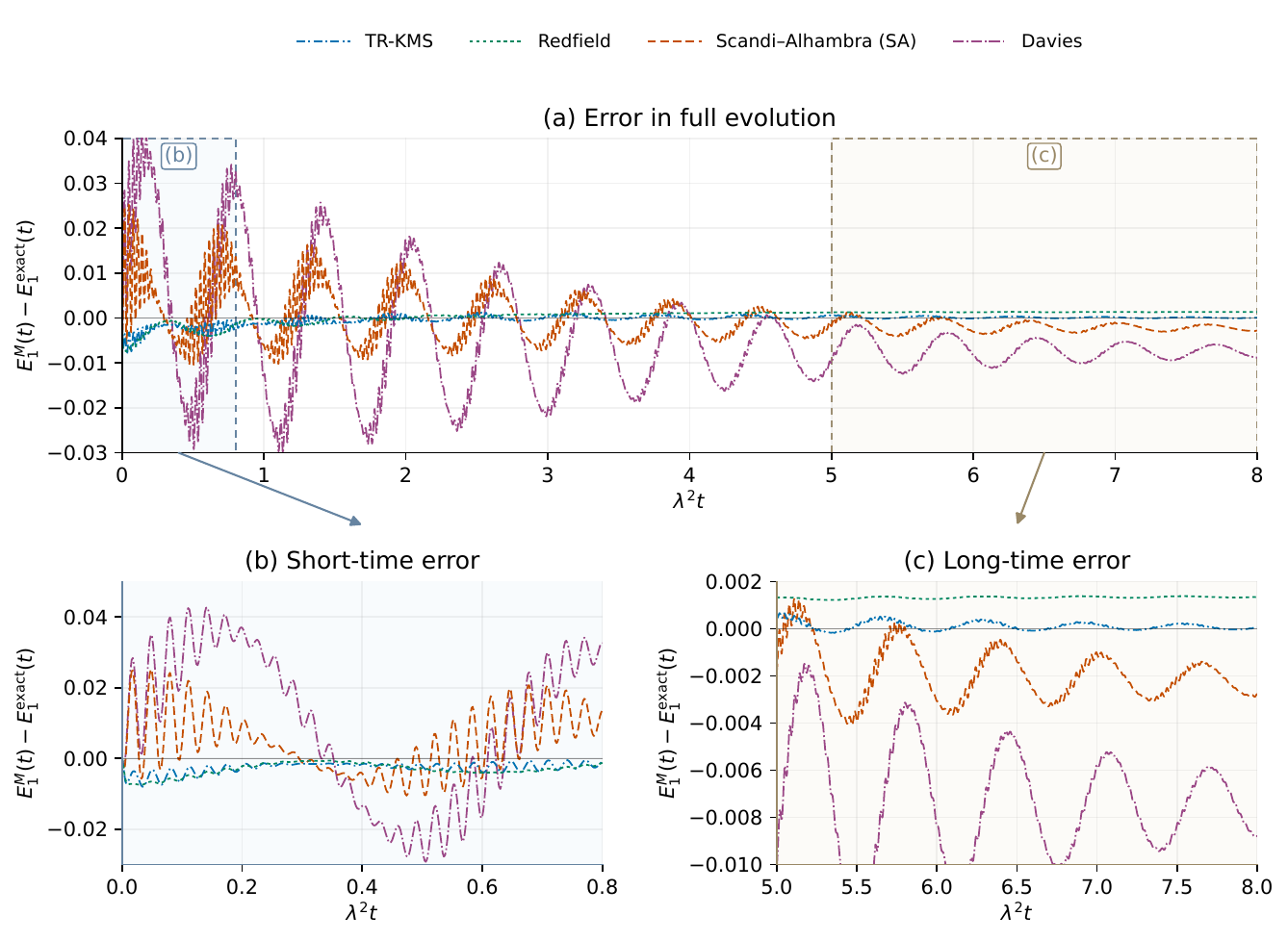}
 \caption{Signed error in the first oscillator's local energy
 relative to exact microscopic dynamics, for $\Delta=\lambda=0.10$
 and the initial state $|\alpha_1=1\rangle\otimes|0\rangle$.
 (a) Evolution over $0\leq\lambda^2t\leq8$.
 The shaded time windows and arrows identify the enlargements in
 (b) the short-time interval $0\leq\lambda^2t\leq0.8$ and
 (c) the late-time interval $5\leq\lambda^2t\leq8$.
 The vertical ranges differ among panels. TR-KMS completion retains a transient
 error close to Redfield while approaching a substantially more
 accurate stationary energy.}
 \label{fig:bench-local-energy}
\end{figure*}
To make the illustration of error comparison clearer, 
we also show how the error of a physical observable evolves for each master equation. 
Specifically, using the same initial state and bath
parameters as above, we investigate the average energy of the first oscillator $E_1^M(t)$ in each master equation, and its error is characterized by the difference from the exact result $E_1^{\rm mic}(t)$,
\begin{equation}\label{eq:bench-local-energy}
 E_1^M(t)=\frac12\langle p_1^2+q_1^2\rangle_M(t),\qquad
 \delta E_1^M(t)=E_1^M(t)-E_1^{\rm mic}(t).
\end{equation}
In Fig.~\ref{fig:bench-local-energy}, this time-dependent energy error is shown for
the Davies, SA, and Redfield equations and TR-KMS completion.
The full evolution in panel (a) shows oscillatory errors whose
amplitudes decrease as the system relaxes. TR-KMS completion closely
follows Redfield during the transient, while SA and Davies exhibit
substantially larger excursions from the exact energy. The
short-time enlargement in panel (b) resolves rapid oscillations
superimposed on the slower variation of the error. Despite these
oscillations, TR-KMS completion and Redfield remain much closer to
zero. The largest deviations are clearly smaller for these two equations.
The late-time enlargement in panel (c) reveals a further distinction.
As the remaining oscillations decay, the TR-KMS error stays close
to zero, whereas Redfield approaches a positive offset and SA and
Davies approach negative offsets. 
TR-KMS completion combines a transient error
close to Redfield with a substantially smaller equilibrium offset,
consistent with the accuracy scalings discussed above.

This analysis justifies the advantage of our master equation: 
Among the detailed-balance master equations compared here, TR-KMS completion dynamics is the only one that retains \(O(\lambda^2)\) dynamical accuracy across all three frequency regimes. This benchmark thus demonstrates that TR-KMS detailed balance can be restored while preserving the second-order accuracy of nonsecular dynamics.

\section{Conclusion}
We have constructed a completely positive quantum master equation that satisfies TR-KMS detailed balance and admits a time-uniform \(O(\lambda^2)\) error bound relative to the microscopic dynamics. 
The equation retains interference between distinct transition frequencies and therefore applies beyond the secular approximation. 
These results show that thermal reversibility and dynamical accuracy can be reconciled without requiring the frequency separation underlying the secular approximation.

The central idea is to complete the Redfield generator while preserving its second-order weak-coupling accuracy. To retain this accuracy, we match both the dissipative and coherent Redfield coefficients at resonance, while imposing detailed balance to restore thermal reversibility. A factorized dissipative kernel permits a representation with at most as many Lindblad operators as microscopic coupling channels. 
The second-order Hamiltonian of mean force serves two
purposes: its Gibbs state agrees with microscopic equilibrium up to
$O(\lambda^4)$, and it makes the Redfield coherent contribution compatible
with TR-KMS resonance matching.

Within our construction, resonance matching and TR-KMS detailed balance strongly constrain the dynamics, with a canonical choice resolving the remaining freedom in the multichannel case. The distinction between TR-KMS and nTR-KMS is essential: nTR-KMS is generically incompatible with the same resonance-matching requirements. Explicit time reversal thus plays a constructive role in reconciling thermal reversibility with microscopic dynamical accuracy.

For a fixed finite-dimensional system, the error bound holds under the assumptions of controlled microscopic Redfield error and relaxation on the dissipative timescale.
The error analysis uses resonance matching to construct an
auxiliary comparison state that differs from the microscopic state by
$O(\lambda^2)$. Its evolution differs from the TR-KMS completion dynamics
by a persistent fourth-order error source, in addition to an initial
transient. Relaxation on the dissipative timescale $O(\lambda^{-2})$
limits the accumulation of this source, yielding a time-uniform
$O(\lambda^2)$ error under the stated conditions.

In an exactly solvable two-oscillator benchmark, we confirm the \(O(\lambda^2)\) dynamical error and $O(\lambda^4)$ equilibrium error across the secular, intermediate, and nonsecular regimes.
Among the detailed-balance equations compared here, TR-KMS completion
is the only one that preserves second-order dynamical accuracy
across all three frequency regimes.

The broader significance of this construction is that detailed balance provides a practical principle for selecting accurate effective dynamics beyond the secular limit. 
Combined with microscopic matching, detailed balance strongly constrains how dissipation and coherent evolution must be modified together. 
These results amount to a systematic microscopic approach of thermodynamically consistent relaxation in open quantum systems.

Our framework opens several directions for further investigation. 
Given the explicit coefficients in the construction, it provides a starting point for studying the interplay between quantum coherence, thermal response, and relaxation in systems with dense spectra. This TR-KMS completion could provide a useful framework for connecting open quantum systems, quantum thermodynamics, and the dynamics of many-body systems.

\section*{Acknowledgements}
J.-H.P. is supported by the National Natural Science Foundation of China (NSFC) (Grant No.
12688201), the Science Challenge
Project (Grant No. TZ2025017).
J.-F.C. acknowledge the support received from the European Union's Horizon Europe research and innovation programme through the ERC StG FINE-TEA-SQUAD (Grant No.~101040729). 
 The views and opinions expressed here are solely those
of the authors and do not necessarily reflect those of the
funding institutions. Neither of the funding institutions
can be held responsible for them. 

\appendix

\section{Time-reversal symmetry and detailed balance}\label{app:db}
This appendix explains the connection between microscopic time reversal,
classical detailed balance, and the quantum KMS detailed balance introduced in the
Preliminary section. 
\subsection{Classical detailed balance}
A classical open system can often be described by a stochastic process.
We denote the system state by $z$. For a system of particles, $z=(r_i,p_i)$
can specify the positions and momenta, while for a chemical reaction
system, $z$ can specify the discrete numbers of molecules of each species. 
We consider time-independent relaxation dynamics with a
unique stationary distribution $\sigma(z)$.

We denote the time-reversed state corresponding to $z$ by
$\mathcal T z$. For phase-space variables, this transformation leaves the
position unchanged and reverses the momentum, $\mathcal T(r_i,p_i)=(r_i,-p_i)$.
For a trajectory $z(t)$ over $0\leq t\leq\tau$, its
time-reversed trajectory is $\bar z(t)=\mathcal T z(\tau-t)$. 
Thus, trajectory reversal combines 
the transformation $\mathcal T$ on each state and the backward evolution of time $t\to\tau-t$. 
In the special case 
where the time-reversal-odd variables, such as momentum, are absent in the dynamics, 
$\mathcal T$ leaves each state unchanged and can be omitted; the reversed trajectory
simply follows the backward evolution.
Otherwise, $\mathcal T$ cannot be thrown away generally.

Detailed balance means that in the stationary state, each forward trajectory and its reverse have the same probability
\begin{equation}
    \Pr_{t\in[0,\tau]}[z(t)] = \Pr_{t\in[0,\tau]}[\bar z(t)].
\end{equation}
Here, $\Pr_{t\in[0,\tau]}[z(t)]$ denotes the trajectory probability density, which is the product of the stationary density of the initial state and the conditional path density.
This reversibility characterizes the dynamical property of
thermal equilibrium. In stochastic thermodynamics, breaking detailed
balance gives rise to a nonzero steady-state entropy production.

For a Markov process, detailed balance can be simply expressed in terms
of transition rates. 
The probability distribution evolves according to a master equation,
\begin{equation}
    \dot\rho(z;t)=\mathcal L\rho (z;t)
       =\int\ud w\,R(z,w)\rho(w;t).
\end{equation}
Here $\mathcal L$ is called the generator, and $R(z,w)$ for $z\neq w$ is the
rate of a transition from $w$ to $z$. The integral notation $\int\ud w$ here represents integration over continuous variables and summation over discrete variables. 
By expressing the trajectory probability in terms of transition rates,
the classical detailed balance condition reads 
\begin{equation}\label{c_db_c}
    R(z,w)\sigma(w)=R(\mathcal T w,\mathcal T z)\sigma(z).
\end{equation}
We have assumed that the steady state is invariant under time reversal, $\sigma(\mathcal Tz)=\sigma(z)$.
Physically, the above relation means that in equilibrium, any transition happens with the same probability as the corresponding reverse transition.

Writing the detailed-balance \eqref{c_db_c} in terms of the generator $\mathcal L$ will help us generalize it to the quantum case.  
We define the transpose of the generator as 
\begin{equation}
    \mathcal{L}^T X (z):=\int\ud w R (w,z)X(w) .
\end{equation}
It is the conjugate operator of $\mathcal L$ under the following inner-product relation $\left(X,Y\right)=\int\ud z X(z) Y(z) $.
We also define a multiplication operator by
\begin{equation}
    M_\sigma X (z) := X(z)\sigma(z),
\end{equation}
which multiplies a function by the steady-state distribution $\sigma$.
Furthermore, the time-reversal operator $\mathcal T$ is generalized to an arbitrary function through $\mathcal{T} X(z) = X(\mathcal{T} z)$.
Now, Eq.~\eqref{c_db_c} is equivalent to the following relation for arbitrary $X(z)$,
\begin{equation*}
    \int \ud w R(z,w)[\sigma(w)X(w)] = \sigma(z) \int\ud w R(\mathcal{T} w,\mathcal{T} z) X(w) ,
\end{equation*}
which gives the classical detailed-balance condition in operator form, 
\begin{equation}\label{c_db_o}
    \mathcal L = M_\sigma \mathcal{T} \mathcal L^T\mathcal{T}^{-1} M_\sigma^{-1}.
\end{equation}
Since the steady state is invariant under $\mathcal{T}$, $\mathcal{T}$ and $M_\sigma$ commute. 
\subsection{Quantum detailed balance}
In open quantum systems, the classical probability distribution is
replaced by a density operator $\rho$. 
A general quantum Markov process can be expressed as 
\begin{equation}
    \dot \rho = \mathcal L \rho,
\end{equation}
with the generator $\mathcal L$. 
We denote the steady-state density operator in Gibbs form $\sigma = \ue^{-\beta H}/Z$,
with Hamiltonian $H$.

The adjoint generator $\mathcal L^\sharp$ is defined through the conjugate relation 
\begin{equation}
    \Tr[X^\dagger \mathcal L(Y)]=\Tr[\mathcal L^\sharp(X)^\dagger  Y],
\end{equation} 
for arbitrary operators $X$ and $Y$. 
It plays a similar role of $\mathcal L^T$ in the classical case.
Furthermore, we introduce the state-reversal superoperator $\mathcal{T}$ acting on the density operator, given by $\mathcal T \rho = \Theta \rho \Theta^{-1}$, 
where $\Theta$ is the antiunitary time-reversal operator acting on pure states. 
As in the classical case, we focus on a time-reversal-invariant Hamiltonian, $\mathcal T H =H$. 

Now, every term in Eq.~\eqref{c_db_o} has a quantum counterpart, except the multiplication operator $M_\sigma$. 
Noncommutativity leads to several possible choices of multiplication operator. 
For example, the following 
\begin{equation}
    M_s \rho :=\sigma^s\rho\sigma^{1-s} ,\quad 0\leq s\leq1,
\end{equation}
has the same classical limit for different $s$.
We focus on the standard case with $s=1/2$ in KMS detailed balance, 
\begin{equation}
    M_\sigma = \sigma^{1/2}\rho\sigma^{1/2}.
\end{equation}
This choice is related to the Petz recovery map \cite{Kwon2022}.

By analogy with Eq.~\eqref{c_db_o}, we obtain the TR-KMS detailed balance condition for a quantum Markov process,
\begin{equation}\label{app:TR-KMS}
    \mathcal L= M_\sigma \mathcal{T} \mathcal L^\sharp\mathcal{T}^{-1} M_\sigma^{-1}.
\end{equation}
As in the classical case, the time-reversal operator $\mathcal T$ should not be omitted in general.

\begingroup
\subsection{TR-KMS relations in the energy representation}
\label{app:db-coefficients}
Here we derive the TR-KMS conditions on the coefficients $K$ and $h$
of the general generator in Eq.~\eqref{generator}. We first compare the
jump terms, which gives the condition on $K$, and then compare the
remaining terms to obtain the relation between $h$ and $K$.

To display these contributions, write the jump term through a superoperator
$\mathcal J_{ij;\omega'\omega}[\rho]
=A_j(\omega)\rho A_i(\omega')^\dagger$ and define two coefficient matrices,
$F_L=K/2+\ui h$ and $F_R=K/2-\ui h$. Then, Eq.~\eqref{generator} reads
\begin{equation}\label{eq:db-app-left-right}
 \begin{aligned}
 \mathcal L[\rho]={}&-\ui[H+N,\rho]+\sum_{ij\omega'\omega}
                   K_{ij}\mathcal J_{ij;\omega'\omega}[\rho]\\
 &-\sum_{ij\omega'\omega}\left[
       (F_L)_{ij}P_{ij;\omega'\omega}\rho
       +(F_R)_{ij}\rho P_{ij;\omega'\omega}\right].
 \end{aligned}
\end{equation}
We denote the thermal reverse of the generator by
\begin{equation}
    \mathcal L^{\rm rev}=M_\sigma\mathcal T\mathcal L^\sharp\mathcal T^{-1}M_\sigma^{-1}.
\end{equation}
In its decomposition as in Eq.~\eqref{eq:db-app-left-right}, we denote the corresponding coefficients by
$K^{\rm rev}$,
$F_L^{\rm rev}$, and
$F_R^{\rm rev}$.
TR-KMS detailed balance states that $\mathcal L^{\rm rev}=\mathcal L$.
We express this condition as relations between the corresponding
coefficients. If the operators in the chosen representation are linearly
dependent, the same generator can have more than one coefficient representation.

We use $\nu=\omega-\omega'$ and the Bohr-operator identities
\begin{equation}\label{eq:db-app-thermal-identities}
 \begin{aligned}
 \sigma^{1/2}A_i(\omega)\sigma^{-1/2}
   &=\ue^{\beta\omega/2}A_i(\omega),\\
 \sigma^{1/2}P_{ij;\omega'\omega}\sigma^{-1/2}
   &=\ue^{\beta\nu/2}P_{ij;\omega'\omega},\\
 \mathcal T A_i(\omega)&=\sum_jE_{ij}A_j(\omega).
 \end{aligned}
\end{equation}
Time reversal preserves the Bohr frequency because $\mathcal T H=H$.
Hermiticity means
$K(\omega',\omega)^\dagger=K(\omega,\omega')$ and
$h(\omega',\omega)^\dagger=h(\omega,\omega')$.

The coherent term $-\ui\operatorname{ad}_{H+N}=-\ui\comm{H+N}{\cdot}$ already satisfies
TR-KMS. Indeed, taking its adjoint reverses the sign of
$\ui$, and time reversal reverses it once more. Since
$\mathcal T(H+N)=H+N$, this gives
\begin{equation}
 \mathcal T(-\ui\operatorname{ad}_{H+N})^\sharp\mathcal T^{-1}
 =-\ui\operatorname{ad}_{H+N}.
\end{equation}
Moreover, $[H,N]=0$ implies that $H+N$ commutes with $\sigma$.
Hence $M_\sigma\operatorname{ad}_{H+N}M_\sigma^{-1}
=\operatorname{ad}_{H+N}$,
so the complete thermal reverse also leaves this term unchanged. We
therefore compare the remaining terms with those of the thermal reverse.

For a jump term, the thermal reverse gives
\begin{equation}
 \begin{aligned}
 &\bigl(M_\sigma\mathcal T
 [K_{ij}(\omega',\omega)\mathcal J_{ij;\omega'\omega}]^\sharp
 \mathcal T^{-1}M_\sigma^{-1}\bigr)[\rho]\\
 &\quad=\ue^{-\beta(\omega+\omega')/2}
 K_{ij}(\omega',\omega)\sum_{ab}E_{ja}E_{ib}\,
 \mathcal J_{ba;-\omega',-\omega}[\rho].
 \end{aligned}
\end{equation}
The exponential follows from the thermal factors of the two Bohr
operators in Eq.~\eqref{eq:db-app-thermal-identities}. Relabelling the
frequencies and channel indices in the full jump sum then gives
\begin{equation}\label{eq:db-app-reversed-jump}
 K^{\rm rev}(\omega',\omega)
 =\ue^{\beta(\omega+\omega')/2}
       E K(-\omega',-\omega)E^T.
\end{equation}
The adjoint conjugates the coefficient, and the antiunitary time
reversal conjugates it once more. Equating $K^{\rm rev}$ to $K$
therefore yields Eq.~\eqref{TR-KMSe1}, without complex conjugation
on its right-hand side.

For the remaining terms, introduce the left multiplication superoperator
$\mathcal P^L_{ij;\omega'\omega}[\rho]=P_{ij;\omega'\omega}\rho$,
with an arbitrary operator $\rho$. 
Direct calculation of its thermal reverse yields
\begin{equation}
 \begin{aligned}
 &M_\sigma\mathcal T\left[(F_L)_{ij}(\omega',\omega)
    \mathcal P^L_{ij;\omega'\omega}\right]^\sharp
    \mathcal T^{-1}M_\sigma^{-1}[\rho]\\
 &\qquad=\ue^{\beta\nu/2}(F_L)_{ij}(\omega',\omega)
    \sum_{ab}E_{ja}E_{ib}\rho P_{ab;\omega\omega'}.
 \end{aligned}
\end{equation}
After the thermal reverse, left multiplication term becomes a right
multiplication term. 
Relabelling the frequencies and channel indices in the full sum gives its coefficient
\begin{equation}
 \begin{aligned}
 F_R^{\rm rev}(\omega',\omega)
 &=\ue^{-\beta\nu/2}E F_L(\omega,\omega')^T E^T
 \end{aligned}
\end{equation}
Similarly, the original right multiplication term gives a left
multiplication term with coefficient
\begin{equation}
 F_L^{\rm rev}(\omega',\omega)
 =\ue^{\beta\nu/2}E F_R(\omega,\omega')^T E^T.
\end{equation}
The TR-KMS detailed balance requires $F_L^\text{rev}=F_L$ and $F_R^\text{rev}=F_R$, 
which gives
\begin{equation}\label{eq:db-app-TR-multiplication}
 \begin{aligned}
 \frac12K+\ui h&=\ue^{\beta\nu/2}\left(\frac12EK^*E^T
                                      +\ui E h^*E^T\right),\\
 \frac12K-\ui h&=\ue^{-\beta\nu/2}\left(\frac12E K^*E^T
                                      -\ui E h^*E^T\right).
 \end{aligned}
\end{equation}
Eliminating $E h^*E^T$ gives, for $\nu\ne0$,
\begin{equation}\label{eq:db-app-TR-h}
 h=-\frac\ui2\coth\!\left(\frac{\beta\nu}{2}\right)K
       +\frac\ui2\operatorname{csch}\!\left(\frac{\beta\nu}{2}\right)EK^*E^T.
\end{equation}
For a parity-preserving kernel, Eq.~\eqref{TR-KMSe1} and Hermiticity give
\begin{equation}\label{eq:hmf-reversed-kernel}
 \ue^{\beta\omega}K(-\omega,-\omega')^T
       =\ue^{\beta\nu/2}E K(\omega',\omega)^*E^T.
\end{equation}
Substituting this identity into Eq.~\eqref{eq:db-app-TR-h} gives
$h(\omega',\omega)=f[K](\omega',\omega)$, with $f[K]$ defined in
Eq.~\eqref{eq:hmf-f}.
Thus, within this coefficient representation, Eqs.~\eqref{TR-KMSe1} and
\eqref{TR-KMSe2} are equivalent to the dissipative reversal and
multiplication coefficient conditions derived above. Substituting Eq.~\eqref{eq:db-app-TR-h} into
Eq.~\eqref{eq:db-app-TR-multiplication} verifies both multiplication
conditions.

At $\nu=0$,  the multiplication conditions~\eqref{eq:db-app-TR-multiplication} instead require
$K=EK^*E^T$ and $h=Eh^*E^T$. The corresponding coherent term commutes
with $H$ and is time-reversal even, giving the freedom represented
by $N$.

\subsection{nTR-KMS relations in the energy representation}
\label{app:db-ntr-coefficients}
For the extended nTR-KMS convention, we need to keep the coherent part $-\ui[H+N,\cdot]$ separate
and apply $M_\sigma(\cdot)^\sharp M_\sigma^{-1}$ to the remaining
generator. Without the second conjugation supplied by time reversal,
the reversed jump coefficient is
\begin{equation}\label{eq:db-app-nTR-K}
 K^{\rm rev}_{\rm nTR}(\omega',\omega)
 =\ue^{\beta(\omega+\omega')/2}K(-\omega',-\omega)^*.
\end{equation}
Equating $K^{\rm rev}_{\rm nTR}$ with $K$ gives the first relation in
Eq.~\eqref{eq:db-ntr-coefficients1}. Since
$K(\omega',\omega)^\dagger=K(\omega,\omega')$, this relation implies
\begin{equation}\label{eq:db-app-ntr-reversed-kernel}
 \ue^{\beta(\omega+\omega')/2}K(-\omega,-\omega')^T=K(\omega',\omega).
\end{equation}
After similar procedures, the left and right multiplication terms give
\begin{equation}\label{eq:db-app-nTR-multiplication}
 \begin{aligned}
 \frac12K+\ui h&=\ue^{\beta\nu/2}\left(\frac12K-\ui h\right),\\
 \frac12K-\ui h&=\ue^{-\beta\nu/2}\left(\frac12K+\ui h\right).
 \end{aligned}
\end{equation}
Solving either equation gives
\begin{equation}\label{eq:db-app-nTR-h}
 h_{\rm nTR}(\omega',\omega)
  =-\frac\ui2\tanh\!\left(\frac{\beta\nu}{4}\right)
                         K(\omega',\omega).
\end{equation}
For $\nu\ne0$, substituting Eq.~\eqref{eq:db-app-ntr-reversed-kernel} into
Eq.~\eqref{eq:hmf-f} gives
\begin{equation*}
 \begin{aligned}
 f[K](\omega',\omega)
 &=\frac{\ui}{2}\bigl[\csch(x)-\coth(x)\bigr]K(\omega',\omega)\\
 &=h_{\rm nTR}(\omega',\omega),
 \end{aligned}
\end{equation*}
where $\coth(x)-\csch(x)=\tanh(x/2)$.
This proves the equivalence of the coherent relations in
Eqs.~\eqref{eq:db-ntr-coefficients2} and \eqref{eq:db-ntr-explicit} under
the nTR-KMS dissipative relation.

\par\endgroup

\section{Expansion of the Hamiltonian of mean force}\label{app:hmf}

The second-order mean-force expansion is established in
Refs.~\cite{TimofeevTrushechkin2022,LobejkoEtAl2024}.
We summarize this expansion and then give the steps
connecting the mean-force correction to the coherent coefficient in this work.

\subsection{Mean-force expansion and equilibrium accuracy}
We first account for the difference between the Redfield stationary state and the exact mean-force state.
To compare them,
we consider a time-local weak-coupling expansion of the microscopic
reduced dynamics,
\begin{equation}
 \dot\rho_{\rm mic}(t)=
 \bigl(\mathcal L_0+\lambda^2\mathcal L_2^R
       +\lambda^4\mathcal L_4+\cdots\bigr)\rho_{\rm mic}(t)
 +\mathcal I(t).
\end{equation}
Here $\mathcal L_0=-\ui\operatorname{ad}_{H_0}$,
$\mathcal L_2^R$ is the Redfield generator, and $\mathcal L_4$ is the
fourth-order contribution in this expansion. The term $\mathcal I(t)$
accounts for the initial transient. Assuming the expansion is regular in the
long-time limit and $\mathcal I(t)\to0$, relaxation to the mean-force
state gives
\begin{equation}
 0=\bigl(\mathcal L_0+\lambda^2\mathcal L_2^R
          +\lambda^4\mathcal L_4+\cdots\bigr)\rho_{\rm MF}.
\end{equation}
We write the expansion of the mean-force state, 
$\rho_{\rm MF}=\rho_0+\lambda^2\rho_2^{\rm MF}
+\lambda^4\rho_4^{\rm MF}+\cdots$, with
$\rho_0=\ue^{-\beta H_0}/Z_0$. Inserting this expansion into the stationary equation and collecting powers of
$\lambda$ gives
\begin{equation}\label{eq:hmf-stationary-orders}
 \begin{aligned}
 0&=\mathcal L_0\rho_2^{\rm MF}+\mathcal L_2^R\rho_0,\\
 0&=\mathcal L_0\rho_4^{\rm MF}
    +\mathcal L_2^R\rho_2^{\rm MF}
    +\mathcal L_4\rho_0.
 \end{aligned}
\end{equation}
Since $(\mathcal L_0 X)_{mn}=-\ui(E_m-E_n)X_{mn}$, the first
equation determines $\rho_2^{\rm MF}$ between different energies,
while leaving its energy-diagonal blocks undetermined. 
Projecting the second equation onto the energy-diagonal blocks gives
\begin{equation}
 0=\mathcal P_0\bigl(\mathcal L_2^R\rho_2^{\rm MF}
                      +\mathcal L_4\rho_0\bigr),
 \qquad
 \mathcal P_0(X)=\sum_\varepsilon\Pi_\varepsilon X\Pi_\varepsilon .
\end{equation}
Thus, determining these diagonal blocks requires the fourth-order contribution
$\mathcal L_4\rho_0$, which is absent from the Redfield
generator~\cite{FlemingCummings2011,ThingnaWangHanggi2012,SubasiEtAl2012}.

The following directly determines the second-order term in Hamiltonian of mean force.
The Hamiltonian of mean force $H^\text{MF}$ is defined from
\begin{equation}\label{eq:hmf-W}
 \ue^{-\beta H^{\rm MF}}=\frac{\Tr_B\ue^{-\beta(H_0+H_B+\lambda V)}}{Z_B}
             .
\end{equation}
Here, $V=\sum_i A_i\otimes B_i$ is the interaction Hamiltonian.
Assuming the thermal expansion exists and odd bath correlations vanish,
the imaginary-time expansion gives~\cite{TimofeevTrushechkin2022,LobejkoEtAl2024}
\begin{equation}\label{eq:hmf-D}
 \begin{aligned}
 \ue^{-\beta H^\text{MF}}&=\ue^{-\beta H_0}[I+\lambda^2D_\beta+O(\lambda^4)],\\
 D_\beta&=\sum_{ij}\int_0^\beta\ud u\int_0^u\ud v\,
 C_{ij}[-\ui(u-v)]A_i(u)A_j(v),
 \end{aligned}
\end{equation}
where $A_i(u)=\ue^{uH_0}A_i\ue^{-uH_0}$ is the imaginary evolution of the coupling channel operator in the interaction picture.
In the first equation, substituting \(H^{\rm MF}=H_0+\lambda^2H_2^{\rm MF}+O(\lambda^4)\) and expanding to second order, we compare the result 
with Eq.~\eqref{eq:hmf-D} to obtain
\begin{equation}\label{eq:hmf-inversion}
 \int_0^\beta\ud u\,\ue^{uH_0}H_2^{\rm MF}\ue^{-uH_0}=-D_\beta.
\end{equation}
In the second equation in Eq.~\eqref{eq:hmf-D}, expanding $A_i(u)$ and $A_j(v)$ in the Bohr operators of $H_0$ gives
and expression for $D_\beta$,
\begin{equation}
 D_\beta=\sum_{ij;\omega'\omega}I_{ij}(\omega',\omega)P^0_{ij;\omega'\omega},
 \qquad P^0_{ij;\omega'\omega}=A_i^0(\omega')^\dagger A_j^0(\omega).
\end{equation}
Writing $\nu=\omega-\omega'$, the coefficients in this expansion are
\begin{equation}\label{eq:hmf-I}
 \begin{aligned}
 I_{ij}(\omega',\omega)
 &=\int_0^\beta\ud u\int_0^u\ud v\,
 C_{ij}[-\ui(u-v)]\ue^{\omega'u-\omega v}\\
 &=\frac{J_{\beta,ij}(\omega')-\ue^{-\beta\nu}J_{\beta,ij}(\omega)}{\nu},
 \end{aligned}
\end{equation}
where $J_\beta(\omega)=\int_0^\beta C(-\ui s)\ue^{\omega s}\ud s$; the second
line follows by setting $s=u-v$ and integrating over $v$.
Applying $[H_0,P^0_{ij;\omega'\omega}]=-\nu P^0_{ij;\omega'\omega}$ to
Eq.~\eqref{eq:hmf-inversion} provides a concrete expression for $H_2^{\rm MF}$,
\begin{equation}\label{eq:hmf-g-derived}
 \begin{aligned}
 H_2^{\rm MF}&=\sum_{ij;\omega'\omega}g_{ij}(\omega',\omega)P^0_{ij;\omega'\omega},\\
 g(\omega',\omega)&=\frac{\ue^{\beta\nu}J_\beta(\omega')-J_\beta(\omega)}
                    {1-\ue^{\beta\nu}}.
 \end{aligned}
\end{equation}
Here $J_\beta(\omega)$ is Hermitian and $g(\omega',\omega)=g(\omega,\omega')$, so $H_2^{\rm MF}$
is Hermitian.

To express this established expansion through the bath Lamb shift,
we use $S(\omega)=\operatorname{PV}\int \ud\mu\,\gamma(\mu)/[2\pi(\omega-\mu)]$
and the thermal relation $\ue^{-\beta\mu}\gamma(\mu)=\gamma(-\mu)^T$.
Then~\cite{LobejkoEtAl2024}
\begin{equation}\label{eq:hmf-J-derived}
 \begin{aligned}
 J_\beta(\omega)&=\int\frac{\ud\mu}{2\pi}\gamma(\mu)
               \frac{\ue^{\beta(\omega-\mu)}-1}{\omega-\mu}\\
           &=-S(\omega)-\ue^{\beta \omega}S(-\omega)^T.
 \end{aligned}
\end{equation}
The two principal-value terms use the same ultraviolet regularization
when one is needed. This proves Eqs.~\eqref{eq:hmf-H2}--\eqref{eq:hmf-J}
in the conventions of the main text.

\subsection{Coherent correction and thermal compensation}
\label{app:hmf-compatibility}
We now make explicit how the mean-force correction reorganizes the
Redfield coherent term. 
With $H^{\rm ref}=H_0+\lambda^2H_2^{\rm MF}$, we rewrite the total Hamiltonian that includes Lamb shift in the Redfield equation as
\begin{equation}
 \begin{aligned}
 H_0+\lambda^2H^\text{LS}&=H^{\rm ref}
       +\lambda^2\sum_{ij;\omega'\omega}h^R_{ij}(\omega',\omega)P^0_{ij;\omega'\omega},\\
 h^R(\omega',\omega)&=r^\text{LS}(\omega',\omega)-g(\omega',\omega).
 \end{aligned}
\end{equation}
Writing $\nu=\omega-\omega'$ and
$z=\ue^{\beta\nu}$, Eqs.~\eqref{eq:hmf-g-derived}
and \eqref{eq:hmf-J-derived} give
\begin{equation}\label{eq:hmf-g-S}
 g(\omega',\omega)=\frac{zS(\omega')-S(\omega)
       +\ue^{\beta \omega}[S(-\omega')^T-S(-\omega)^T]}{z-1}.
\end{equation}
The Redfield coherent coefficient $r^\text{LS}(\omega',\omega)$ is given in Eq.~\eqref{eq:prelim-Redfield-coefficients},
\begin{equation}
    r^\text{LS}(\omega',\omega)
       =\frac12[S(\omega)+S(\omega')]
                       -\frac\ui4[\gamma(\omega)-\gamma(\omega')].
\end{equation}
For $\omega'\ne\omega$, direct subtraction gives the coefficient in the main text:
\begin{equation}\label{eq:hmf-hR-derived}
 \begin{aligned}
 &h^R(\omega',\omega)\\={}&\frac{(1+z)[S(\omega)-S(\omega')]
             -2\ue^{\beta \omega}[S(-\omega')^T-S(-\omega)^T]}{2(z-1)}\\
 &-\frac\ui4[\gamma(\omega)-\gamma(\omega')].
 \end{aligned}
\end{equation}
Using the thermal compensation functional defined in Eq.~\eqref{eq:hmf-f},
insert $K^R(\omega',\omega)=[\gamma(\omega')+\gamma(\omega)]/2+\ui[S(\omega)-S(\omega')]$.
Expanding the functional and using the thermal relation for $\gamma$ gives
\begin{equation}\label{eq:hmf-f-eval}
 \begin{aligned}
 &f[K^R](\omega',\omega)=-\frac{\ue^{\beta \omega}[S(-\omega')^T-S(-\omega)^T]}{z-1}\\
 &+\frac{1+z}{2(z-1)}[S(\omega)-S(\omega')]
       -\frac\ui4[\gamma(\omega)-\gamma(\omega')]\\
 &={}h^R(\omega',\omega).
 \end{aligned}
\end{equation}
The above gives Eq.~\eqref{eq:reference-hR} in the main text. 
For smooth bath coefficients, the continuous resonance limits of $\omega^\prime\to\omega$ can be 
calculated directly
\begin{equation}\label{eq:hmf-hR-diagonal}
 \begin{aligned}
 g(\omega,\omega)&=S(\omega)-\frac{S'(\omega)-\ue^{\beta \omega}S'(-\omega)^T}{\beta},\\
 h^R(\omega,\omega)&=\frac{S'(\omega)-\ue^{\beta \omega}S'(-\omega)^T}{\beta}.
 \end{aligned}
\end{equation}
Here $S'(-\omega)$ denotes the derivative evaluated at $-\omega$. In the
single-channel completion, the last expression is
$2\gamma(\omega)\phi'(\omega)/\beta$, giving the coherent matching condition
in Sec.~\ref{sec:completion}.

\section{Derivation and properties of the multichannel completion}
\label{app:completion-details}
This appendix supplies the steps behind the multichannel phase
equation and the canonical choice in Sec.~\ref{sec:completion}.

\subsection{Continuous resonance limit}
We drive the continuous limit of
$h^K(\omega',\omega)$ in
Eq.~\eqref{eq:multi-completion-coefficients} as $\omega'\to\omega$.
At fixed $\omega$, set $\omega'=\omega-\nu$. Writing the definition
of $f[K]$ in Eq.~\eqref{eq:hmf-f} in terms of exponentials and using the
dissipative TR-KMS identity~\eqref{eq:hmf-reversed-kernel}, we obtain
\begin{equation}
\begin{aligned}
 h^K(\omega',\omega)
 =\frac{\ui}{\ue^{\beta\nu}-1}\Bigl[
 &\ue^{\beta\nu/2}E K^K(\omega',\omega)^*E^T\\
 &-\frac{1+\ue^{\beta\nu}}{2}K^K(\omega',\omega)\Bigr].
\end{aligned}
\end{equation}
Both thermal prefactors in the numerator are
$1+\beta\nu/2+O(\nu^2)$. At resonance,
$K^K(\omega,\omega)=\gamma(\omega)$ and
$E\gamma(\omega)^*E^T=\gamma(\omega)$, so the numerator vanishes
at $\nu=0$. Since $\ue^{\beta\nu}-1=\beta\nu+O(\nu^2)$, the limit is
set by the first-order difference of the two kernels,
\begin{equation}
 h^K(\omega,\omega)
 =\frac{\ui}{\beta}\lim_{\nu\to0}
 \frac{E K^K(\omega-\nu,\omega)^*E^T-K^K(\omega-\nu,\omega)}{\nu}.
\end{equation}

The factorization $K^K(\omega',\omega)=Q(\omega')^\dagger Q(\omega)$
turns this first-order expansion into a derivative of $Q=UR$, with
$R=\sqrt{\gamma}$ and $U$ unitary. We write
\begin{equation}\label{eq:completion-app-Q-derivative}
 \begin{aligned}
 Q^\dagger Q'&=RU^\dagger U'R+RR'
                     =\frac12\gamma'+\Xi,\\
 \Xi&=RU^\dagger U'R+\frac12[R,R'],
       \qquad \Xi^\dagger=-\Xi.
 \end{aligned}
\end{equation}
Here $\gamma'=R'R+RR'$, and unitarity makes $U^\dagger U'$
anti-Hermitian. Expanding the the kernel in order of $\mu$ gives
\begin{equation}\label{eq:completion-app-K-expansion}
 \begin{aligned}
 K^K(\omega-\nu,\omega)
 &=\gamma-\nu Q'^\dagger Q+O(\nu^2)\\
 &=\gamma-\nu\left(\frac12\gamma'-\Xi\right)+O(\nu^2).
 \end{aligned}
\end{equation}
Since $E\gamma'^*E^T=\gamma'$ and $\Xi^*=-\Xi^T$, the difference
entering the limit is
\begin{equation}\label{eq:completion-app-resonant-numerator}
 \begin{aligned}
 &E K^K(\omega-\nu,\omega)^*E^T-K^K(\omega-\nu,\omega)\\
 &\qquad=-\nu\left(\Xi+E\Xi^TE^T\right)+O(\nu^2).
 \end{aligned}
\end{equation}
It follows that
\begin{equation}\label{eq:completion-app-resonant-h}
 h^K(\omega,\omega)
   =-\frac{\ui}{\beta}\left(\Xi+E\Xi^TE^T\right),
\end{equation}
which produces Eq.~\eqref{eq:continuous-hK} in the main text.

\subsection{Symmetry compatibility and unitary freedom}\label{app:completion-symmetry}
We show that the unitary matrix $U$ obtained from phase equation preserves is compatible with Eq.~\eqref{eq:multi-U-symmetry}.

The thermal and time-reversal symmetries of the bath correlations,
together with Eqs.~\eqref{eq:prelim-bath-transform} and
\eqref{gamma_r}, give
\begin{equation}\label{eq:completion-app-bath-symmetries}
 \begin{aligned}
 &ES(\omega)^TE^T=S(\omega),\quad ED(\omega)^TE^T=D(\omega),\\
 &R(\omega)=\ue^{\beta\omega/2}ER(-\omega)E^T,\quad
 D(\omega)=-\ue^{\beta\omega}ED(-\omega)E^T.
 \end{aligned}
\end{equation}
Differentiating the relation for $R$ yields
\begin{equation}\label{eq:completion-app-R-derivative-symmetry}
 \begin{aligned}
 R'(\omega)&=\frac\beta2R(\omega)
             -\ue^{\beta\omega/2}ER'(-\omega)E^T,\\
 [R,R'](\omega)&=-\ue^{\beta\omega}E[R,R'](-\omega)E^T.
 \end{aligned}
\end{equation}
We recall the phase equation 
\begin{equation}\label{eq:completion-app-phase-connection}
 U^\dagger U^\prime=R^{-1}\left(\frac\ui2D+Y-\frac12[R,R']\right)R^{-1}.
\end{equation}
If we have
\begin{equation}\label{eq:completion-app-Y-frequency-symmetry}
 Y(\omega)=-\ue^{\beta\omega}EY(-\omega)E^T,
\end{equation}
then Eqs.~\eqref{eq:completion-app-bath-symmetries}-
\eqref{eq:completion-app-phase-connection} give
\begin{equation}
    U(\omega)^\dagger U^\prime(\omega)=-EU^\dagger(-\omega)U^\prime(-\omega)E^T.
\end{equation}
Consequently, matrix $EU(-\omega)E^T$ solves the same differential equation~\eqref{eq:completion-app-phase-connection}
as $U(\omega)$. Both of them equal $I$ at $\omega=0$, so they are equal to each other,
\begin{equation}\label{eq:completion-app-frequency-reversal}
 \begin{aligned}
 U(\omega)&=EU(-\omega)E^T.
 \end{aligned}
\end{equation}
This proves the required compatibility.

\subsection{Independent baths and minimum deviation}\label{app:canonical choice}
We examine two properties motivating the choice $Y=0$: it preserves the additivity of independent baths and minimizes a normalized near-resonant deviation from the Redfield dissipative kernel.

Consider mutually independent single-channel baths, with the same
reference Hamiltonian for each contribution. Additivity requires the
multichannel completion to equal the sum of the corresponding single-channel contributions. 
In this case, $\gamma$ and $S$ are
diagonal, and so are $R$ and $D$. 
Each channel has a time-reversal parity
$E_{ii}=\pm1$, so $EY^TE^T=-Y$ requires $Y_{ii}=0$. For $i\ne j$,
Eq.~\eqref{eq:completion-app-K-expansion} gives
\begin{equation}\label{eq:completion-app-independent-slope}
 K^K_{ij}(\omega-\nu,\omega)
       =\nu Y_{ij}(\omega)+O(\nu^2).
\end{equation}
A nonzero $Y_{ij}$ therefore introduces a cross-channel dissipative
term between independent baths. Additivity requires these terms to
vanish, fixing $Y=0$. With this choice and $U(0)=I$, the matrices
$U$ and $Q$ are diagonal and each contribution retains its
single-channel form.

We next compare the our constructed dissipative coefficient $K^K(\omega^\prime,\omega)$ and the Redfield coefficient $K^R(\omega^\prime,\omega)$ near
resonance. Their values agree at $\omega'=\omega$, and the difference starts at first order in $\nu=\omega-\omega^\prime$. 
We use the following matrix to characterize their difference, which has been normalized by $R^{-1}=\gamma^{-1/2}$ on both sides,
\begin{equation}\label{eq:completion-app-normalized-deviation}
 \begin{aligned}
 \delta_Y(\omega)
 &:=\lim_{\nu\to0}R^{-1}
    \frac{1}{\nu} [K^K(\omega-\nu,\omega)-K^R(\omega-\nu,\omega)]
    R^{-1}\\
 &=\ui R^{-1}\left(\frac12D-S'\right)R^{-1}+R^{-1}YR^{-1}.
 \end{aligned}
\end{equation}
We quantify the normalized deviation by the 
Frobenius norm $\norm{\delta_Y}_2^2=\Tr(\delta_Y^\dagger\delta_Y)$. 
Write
$\delta_Y=\Delta_0+B_Y$, where $B_Y=R^{-1}YR^{-1}$. 
The two terms have different symmetries,
\begin{equation}
 E \Delta_0 E^T=\Delta_0,\qquad
 E B_Y E^T=-B_Y.
\end{equation}
The operation $E\cdot E^T $ preserves the matrix
inner product $\left( A,B\right)_{\rm F}=\Tr(A^\dagger B)$. 
It implies that the two terms are orthogonal, 
\begin{equation}
 \left(\Delta_0,B_Y\right)
 =\left(E \Delta_0 E^T, E B_YE^T\right)
 =-\left(\Delta_0,B_Y\right)=0.
\end{equation}
Thus, when taking the norm of $\delta_Y$, 
the cross term vanishes, giving
\begin{equation}\label{eq:completion-app-minimum-deviation}
 \norm{\delta_Y}_2^2
 =\norm{\Delta_0}_{\rm F}^2+\norm{R^{-1}YR^{-1}}_2^2.
\end{equation}
Thus, the choice of $Y=0$ uniquely minimizes $\norm{\delta_Y}_2$ at each frequency, which quantifies the normalized
first-order dissipative 
from Redfield equation.

\bibliography{KSM-completion}
\end{document}